\def \RemoveRedundantBCARET {}

\def \KeepNotesComments {}
	\def \NotNeedToReducePages {}

\documentclass[runningheads,a4paper]{llncs}

\usepackage{float}

\usepackage[utf8]{inputenc}
\usepackage{verbatim}
\usepackage{enumitem}
\usepackage{ marvosym }

\usepackage{amssymb}
\usepackage{graphicx}
\usepackage{multirow}
\usepackage{ mathrsfs }

\usepackage{listings}

\usepackage{mathtools}

\usepackage[utf8]{inputenc}
\usepackage[english]{babel}
\usepackage{ upgreek }
\usepackage{ amssymb }
\usepackage{ stmaryrd }
\usepackage{graphicx}
\usepackage{amsmath}
\usepackage{amssymb}

\usepackage{fancyhdr}

\usepackage{amsmath}

\usepackage{showexpl}

\usepackage[usenames, dvipsnames]{color} 

\lstdefinestyle{numbers}
{numbers=left, stepnumber=1, numberstyle=\tiny, numbersep=10pt}
\lstdefinestyle{nonumbers}
{numbers=none}

\graphicspath{ {images/} }

\newtheorem{theorem-non}{}

\newcounter{mylabelCounter}

\newcounter{myIntuitionCounter}
\newcommand{\myIntuitionLabel}[1]{\refstepcounter{myIntuitionCounter} (A\themyIntuitionCounter)  \label{#1}}
\newcommand{\refMyIntuitionLabel}[1]{(A\ref{#1})}

\newcommand{\changeallfullversiona}[1]{#1} 
\newcommand{\changeallfullversionb}[1]{#1} 
\newcommand{\changeallfullversionc}[1]{#1} 
\newcommand{\changeallfullversiond}[1]{#1} 
\newcommand{\changeallfullversione}[1]{#1} 

\newcommand{\changeallfullversionblue}[1]{#1} 

\newcommand{\changeifma}[1]{#1} 

\newenvironment{mycolorforparablue}{
}{%
}%

\newenvironment{mycolorforparaallfullversion}{
	\leavevmode\color{red}\ignorespaces%
}{%
}%

\ifdefined \UseFullVersion 

\else

\fi

\ifdefined \KeepNotesComments 
\newcommand{\notes}[1]{} 
\newcommand{\todoh}[1]{} 

\newcommand{\todohide}[1]{} 
\newcommand{\notesx}[1]{} 
\newcommand{\notesy}[1]{} 
\newcommand{\notesh}[1]{} 
\newcommand{\removed}[1]{} 
\newcommand{\done}[1]{} 

\newcommand{\todo}[1]{\bigskip \textcolor{orange}{TODO: #1} \medskip} 

\newcommand{\todon}[1]{\medskip \noindent \textcolor{orange}{NOTES (WILL BE REMOVED LATER): #1} \medskip} 

\newcommand{\notesnta}[1]{\medskip \noindent \textcolor{VioletRed}{NOTES (WILL BE REMOVED LATER): #1} \medskip} 

\else 
\newcommand{\willdelete}[1]{} 
\newcommand{\notes}[1]{} 
\newcommand{\todo}[1]{} 
\newcommand{\todoh}[1]{} 

\newcommand{\todohide}[1]{} 
\newcommand{\notesx}[1]{} 
\newcommand{\notesy}[1]{} 
\newcommand{\notesh}[1]{} 
\newcommand{\removed}[1]{} 
\newcommand{\done}[1]{} 

\newcommand{\notesnta}[1]{} 
\newcommand{\todon}[1]{} 
\fi

\ifdefined \RemoveRedundantBCARET 

\newcommand{\notesctl}[1]{} 

\newcommand{\changenaa}[1]{#1} 
\newcommand{\changenaab}[1]{#1} 
\newcommand{\changenaac}[1]{#1} 
\newcommand{\changenaad}[1]{#1} 
\newcommand{\changenaae}[1]{#1} 
\newcommand{\changenaaf}[1]{#1} 
\newcommand{\changenaag}[1]{#1} 
\newcommand{\changenaaj}[1]{#1} 
\newcommand{\changenaal}[1]{#1} 
\newcommand{\changenaam}[1]{#1} 

\newcommand{\changenaan}[1]{#1} %
\newcommand{\changenaao}[1]{#1} %
\newcommand{\changenaap}[1]{#1} %
\newcommand{\changenaaq}[1]{#1} %
\newcommand{\changenaar}[1]{#1} %

\newcommand{\changenaas}[1]{#1} 
\newcommand{\changenaasbb}[1]{#1} 

\newcommand{\changenaat}[1]{#1} 

\newcommand{\changenaau}[1]{#1} 
\newcommand{\changenaav}[1]{#1} 

\newcommand{\changenaaw}[1]{#1} 

\else 

\newcommand{\changenaa}[1]{#1} 
\newcommand{\changenaab}[1]{#1} 
\newcommand{\changenaac}[1]{#1} 
\newcommand{\changenaad}[1]{#1} 
\newcommand{\changenaae}[1]{#1} 
\newcommand{\changenaaf}[1]{#1} 
\newcommand{\changenaag}[1]{#1} 
\newcommand{\changenaaj}[1]{#1} 
\newcommand{\changenaal}[1]{#1} 
\newcommand{\changenaam}[1]{#1} 

\newcommand{\changenaan}[1]{#1} %
\newcommand{\changenaao}[1]{#1} %
\newcommand{\changenaap}[1]{#1} %
\newcommand{\changenaaq}[1]{#1} %
\newcommand{\changenaar}[1]{#1} %

\newcommand{\changenaas}[1]{\textcolor{red}{#1}} 
\newcommand{\changenaasbb}[1]{\textcolor{red}{#1}} 
\newcommand{\changenaat}[1]{\textcolor{blue}{#1}} 

\newcommand{\changenaau}[1]{\textcolor{blue}{#1}} 
\newcommand{\changenaav}[1]{\textcolor{red}{#1}} 

\newcommand{\changenaaw}[1]{\textcolor{red}{#1}} 

\newcommand{\notesctl}[1]{\bigskip \textcolor{WildStrawberry}{NOTES (WILL BE REMOVED LATER): #1} \medskip} 

\fi

\newcommand{\changeni}[1]{\textcolor{red}{#1}} 

\newcommand{\changeza}[1]{#1} 

\ifdefined \DifferentColorForTextToModify 

\newcommand{\change}[1]{#1} 

\newenvironment{mycolorforparas}{
	\leavevmode\color{red}\ignorespaces%
}{%
}%

\newenvironment{myColorForRemovePages}{
	
	\leavevmode\color{orange}\ignorespaces%
}{%
}%

\ifdefined \OneColorForTheLastChange 

\newcommand{\changenta}[1]{\textcolor{blue}{#1}} 

\else

\newcommand{\changenta}[1]{\textcolor{red}{#1}} 

\fi

\else

\newcommand{\change}[1]{#1} 

\newcommand{\changenta}[1]{#1} 

\fi

\usepackage[svgnames]{xcolor}
\definecolor{darkgreen}{rgb}{0, 0.5, 0}
\definecolor{darkpurple}{rgb}{0.7, 0, 0.7}
\definecolor{darkblue}{rgb}{0, 0, 0.7}

\usepackage{graphicx}
\usepackage{caption}
\usepackage{subcaption}

\usepackage{tikz}
\usetikzlibrary{shapes,arrows}
\usetikzlibrary{shapes,positioning}

\tikzstyle{block} = [draw, fill=blue!20, rectangle, 
minimum height=3em, minimum width=6em]
\tikzstyle{sum} = [draw, fill=blue!20, circle, node distance=1cm]
\tikzstyle{input} = [coordinate]
\tikzstyle{output} = [coordinate]
\tikzstyle{pinstyle} = [pin edge={to-,thin,black}]

\tikzset{myellipse/.style={ellipse,draw,minimum width=1.3cm,minimum height=5mm},
	myrectangle/.style={draw,text width=1.2cm,align=center,midway,font=\scriptsize},
	branode/.style={midway,font=\scriptsize}}

\usepackage{amsthm}
\newtheoremstyle{exampstyle}
{0.3} 
{0.3} 
{} 
{} 
{\bfseries} 
{.} 
{.1em} 
{} 
\theoremstyle{exampstyle} 
\theoremstyle{exampstyle} 
\theoremstyle{exampstyle} 

\usepackage[
colorlinks=true,
\ifdefined\VersionWithComments
pagebackref=true,
\fi
citecolor=darkgreen,
linkcolor=darkblue,
urlcolor=darkpurple,
]{hyperref}

\ifdefined \VersionWithComments
\usepackage{marginnote}
\newcommand{\marginX}{\marginnote{\huge{\quad\quad\textbf{!}\quad\quad}}}
\newcommand{\ea}[1]{{\color{violet}\marginX{}\textbf{[\'Etienne}: #1]}}
\newcommand{\hv}[1]{{\color{green!50!black}\marginX{}\textbf{[Huu-Vu}: #1]}}
\newcommand{\lp}[1]{{\color{magenta}\marginX{}\textbf{[Laure}: #1]}}
\else
\newcommand{\ea}[1]{}
\newcommand{\hv}[1]{}
\newcommand{\lp}[1]{}
\fi

\begin{document}
	
	\mainmatter 
	
	\title{Branching Temporal Logic of Calls and Returns for Pushdown Systems}
	
	\addtolength{\textheight}{0.1cm}
	
	\titlerunning{Branching Temporal Logic of Calls and Returns for Pushdown Systems}
	
	%
	%
	\author{Huu-Vu Nguyen$^1$, Tayssir Touili$^2$}
	\authorrunning{}
	
	\institute{$^1$ \changenaat{LIPN, CNRS and University Paris 13, France} \\ $^2$ \changenta{CNRS, LIPN  and University Paris 13, France}}

	%
	%
	
	\maketitle
	
	\ifdefined \VersionWithComments
	\textcolor{red}{\textbf{This is the version with comments. To disable comments, comment out line~3 in the \LaTeX{} source.}}
	\fi

	\lp{
		Some advice on writing tool papers can be found here: \url{http://www.informatik.uni-hamburg.de/TGI/PetriNets/sc-info/docs/ToolFormat.pdf}}
	
	\ea{Guidelines: 
		
		\textbf{
			Tools papers of a maximum of 4 pages should describe an operational tool and its contributions; 2 additional pages of appendices are allowed that will not be included in the proceedings. Tool papers should explain enhancements made compared to previously published work. A tool paper need not present the theory behind the tool but can focus more on its features, and how it is used, with screen shots and examples. Authors of tools papers should make their tool available for use by reviewers.
		}
	}
	
	\begin{abstract}
		

		Pushdown Systems (PDSs) are a natural model for sequential programs with (recursive) procedure calls. In this work, we define \changenaap{the Branching temporal logic of CAlls and RETurns (BCARET)} that allows to write branching temporal formulas while taking into account the matching between calls and returns. \changenaap{We consider} the model-checking problem of PDSs against BCARET formulas with \changenaaj{"standard"} valuations \changenaaj{(where an atomic proposition holds at a configuration $c$ or not depends only on the control state of $c$, not on its stack)} as well as regular valuations \changenaaj{(where  the set of configurations in which an atomic proposition holds \changenaaq{is regular)}}. We show that these problems can be effectively solved by a reduction to the emptiness problem of Alternating B\"{u}chi Pushdown Systems. \changenaap{We show that our results} can \changenaaj{be applied} \changenaas{for malware detection.}
	\end{abstract}
	
	
	\section{Introduction}


Pushdown Systems (PDSs) are a natural model for sequential programs with (recursive) procedure calls. Thus, it is very important to have model-checking algorithms \changenaan{for} PDSs. A lot of \changenaan{work} focuses on proposing verification algorithms for PDSs, e.g, for both linear temporal logic (LTL and its extensions) \cite{DBLP:conf/concur/BouajjaniEM97,DBLP:journals/iandc/EsparzaKS03,DBLP:conf/cav/EsparzaHRS00,DBLP:journals/entcs/FinkelWW97,DBLP:conf/birthday/KupfermanPV10,DBLP:conf/tacas/SongT13} and branching temporal logic \changeallfullversionb{(CTL and its extensions)} \cite{DBLP:conf/concur/BouajjaniEM97,DBLP:conf/vmcai/Bozzelli06,DBLP:conf/icalp/BurkartS97,DBLP:conf/cav/Walukiewicz96,DBLP:conf/concur/SongT11}. \changeallfullversionb{However, LTL and CTL are not always adequate to specify properties}. \changenaat{Indeed, \changenaau{some properties} need to talk about \changenaau{matching between calls and returns.}} \changenaat{Thus, CARET (a temporal logic of calls and returns) was introduced by Alur et al \cite{DBLP:conf/tacas/AlurEM04}}. This logic allows to write linear temporal logic formulas while taking into account matching of calls and returns. \changenaat{Later, VP-$\mu$  (also named NT-$\mu$ in other works of the same \changenaau{authors}) \cite{DBLP:conf/popl/AlurCM06,DBLP:conf/cav/AlurCM06,DBLP:journals/toplas/AlurCM11}, a branching-time temporal logic that allows to talk about matching between calls and returns, was introduced.} \changenaau{VP-$\mu$ can be seen as an extension of the \changeallfullversione{modal $\mu$-calculus} which allows to talk about \changenaau{matching of calls and returns}.}


In \changenaat{\cite{DBLP:conf/popl/AlurCM06}, the authors proposed an algorithm to model-check VP-$\mu$ formulas for Recursive State Machines (RSMs) \cite{DBLP:journals/toplas/AlurBEGRY05}.} RSMs can be seen as a natural model to \changenaau{represent sequential programs} with \changenaau{(recursive)} procedure calls. Each procedure is modelled as a module. The invocation to a procedure is modelled as a \textit{call} node; the return from a module corresponds to a $ret$ node; and the remaining statements are considered as internal nodes in the RSMs. 	Thus, RSMs are a good formalism to model sequential programs written in \change{structured} programming languages like C or Java. However, they become non suitable for modelling binary or assembly programs; since, in these programs, explicit push and pop instructions can occur. This makes impossible the use of RSMs to model assembly programs and binary codes directly \changenaau{(whereas Pushdown Systems can model binary codes in a natural way \cite{DBLP:conf/fm/SongT12})}. Model checking binary and assembly programs is very important. Indeed, sometimes, only the binary code is available. Moreover, malicious programs are often executables, i.e., binary codes. \changenaat{Thus, it is very important to be able to model check binary and assembly programs against \changenaat{branching-time formulas with matchings between calls and returns}}. One can argue that from a binary/assembly program, one can compute a PDS as described in \cite{DBLP:conf/fm/SongT12} and then apply the translation in \cite{DBLP:journals/toplas/AlurBEGRY05} \changenaat{to obtain a RSM and then apply the VP-$\mu$ model-checking algorithm of \cite{DBLP:conf/popl/AlurCM06} on this RSM}. However, by doing so, we loose the explicit manipulation of the program's stack. Explicit push and pop instructions are not represented in a natural way anymore, and the stack of the RSM does not correspond to the stack of the assembly program anymore. Thus, it is not possible to state intuitive formulas that correspond to properties of the program's behaviors on the obtained RSM. Especially, when these formulas talk about the content of the program's stack. Thus, it is very important to have a \textit{direct} algorithm \changenaat{for model-checking a branching-time temporal logic with matching of calls and returns for} PDSs.

However, VP-$\mu$ is \changenaav{a heavy formalism that can't be used by novice users.} \changenaau{Indeed},  VP-$\mu$ can be \changenaau{seen} as an extension of the modal $\mu$ calculus \changenaau{with several modalities} $\langle loc \rangle$,  $[ loc ]$, $\langle call \rangle$,  $[ call ]$, $\langle ret \rangle$,  $[ ret ]$ \changenaat{that allow} to distinguish between calls, returns, \changenaav{and} other statements (neither calls nor returns). \changenaav{Writing a simple specification in VP-$\mu$ \changenaav{is complicated}. For example, the following \changenaav{simple} property stating that "\changenaav{the configuration $e$} can be reached in the same procedural context as the current configuration" can be described (as \changenaav{shown} in \cite{DBLP:conf/popl/AlurCM06}) by the \changenaav{complex} VP-$\mu$ formula 	 
	$\varphi'_2 = \mu X(e \vee \langle loc \rangle X \vee \langle call  \rangle  \varphi'_3 \{ X\})$  where $\varphi'_3 = \mu Y (\langle ret \rangle R_1 \vee \langle loc \rangle Y \vee \langle call \rangle Y \{Y\} \changenaaw{)}$. } Thus, we need to define a more intuitive branching-time temporal logic (in the style of CTL) that allow to talk naturally and intuitively about matching calls and returns.

\changenaaq{Therefore, we define in this work the \changenaau{Branching temporal logic of CAlls and RETurns BCARET}}. \changenaau{BCARET can be seen as an extension of CTL with operators \changenaav{that allow} to talk about matchings between calls and returns}. \changenaav{Using BCARET, the above reachability property can be described in a simple way by the formula $EF^a e$ where $EF^a$ is a BCARET operator that means "there exists a \changeifma{run} on which eventually in the future in the same procedural context".} \changenaau{We} consider the model-checking problem of PDSs against BCARET formulas with "standard" valuations (where an atomic proposition holds at a configuration $c$ or not depends only on the control state of $c$, not on its stack) as well as regular valuations (where the set of configurations in which an atomic proposition holds is a regular set of configurations). We show that these problems can be effectively solved by a reduction to the emptiness problem of Alternating B\"{u}chi Pushdown Systems (ABPDSs). The latter problem can be solved effectively in \cite{DBLP:conf/concur/SongT11}. \changenaau{Note that the regular valuation case cannot be solved by translating the PDSs to RSMs since as said previously, by doing the translation of PDSs to obtain RSMs, we loose the structure of the program's stack.}


The rest of \changenaas{the paper} is organized as follows. In Section 2, we define Labelled Pushdown Systems. In Section 3, we define the logic BCARET. Section 4 presents \changenaas{applications of BCARET in specifying malicious behaviours}. \changenaap{Our} algorithm to reduce BCARET model-checking to the membership problem of ABPDSs is presented in Section 5. \changeifma{Section 6} \changenaap{discusses} the model-checking problem for PDSs against BCARET formulas with regular valuations. \changeifma{Finally, we conclude in Section \ref{BCARET_sec:conclusion}.}


\label{BCARET_sec:pushdownSystem}
\section{Pushdown Systems: A model for sequential programs}
\label{BCARET_sec:PDS}
Pushdown systems is a \changenaad{natural} model that was extensively used to model sequential programs. Translations from sequential programs to PDSs can be found e.g. in \cite{schwoonThesis}. \changenaab{As will be discussed in the next section, to precisely describe malicious behaviors as well as context-related properties, we need to keep track of the call and return actions in each path. Thus, \changenaal{as done in \cite{DBLP:conf/SAC/VuTayssir2017},} we adapt the PDS model in order to record whether a rule of a PDS corresponds to a {\em call}, a {\em return}, or another instruction. We call this model a {\em Labelled Pushdown System}. We also extend the notion of {\em \changenaad{\changeifma{run}}} in order to take into account matching returns of calls.}

\begin{definition}
	A Labelled Pushdown System (PDS) $\mathcal{P}$ is a tuple $(P, \Gamma, \Delta, \changenaa{\sharp})$, where \changenaal{$P$} is a finite set of control locations, $\Gamma$ is a finite set of stack alphabet, \changenaa{\changenaad{$\sharp \notin \Gamma$} is a bottom stack symbol} and $\Delta$ is a finite subset of
	$((P \times \Gamma) \times (P \times \Gamma^{*}) \times \{call, ret, int \})$. If $((p, \gamma), (q, \omega), t) \in \Delta$ ($t \in \{call, ret, int \}$), we also write $\langle p, \gamma \rangle \xrightarrow{t} \langle q, \omega \rangle \in \Delta$. Rules of $\Delta$ are of the following form, where $p\in P, q\in P, \gamma, \gamma_1, \gamma_2 \in\Gamma$, and $\omega\in\Gamma^*$:
	\begin{itemize}
		\item{($r_1$): $\langle p, \gamma \rangle \xrightarrow{call} \langle q, \gamma_1 \gamma_2 \rangle$}
		\item{($r_2$): $\langle p, \gamma \rangle \xrightarrow{ret} \langle q, \epsilon\rangle$}
		\item{($r_3$): $\langle p, \gamma \rangle \xrightarrow{int} \langle q, \omega\rangle$}
	\end{itemize}
\end{definition}
Intuitively, a rule of the form $\langle p, \gamma \rangle \xrightarrow{call} \langle q, \gamma_1\gamma_2\rangle$ corresponds to a call statement. Such a rule usually models a statement of the form $\gamma \xrightarrow{call~~proc} \gamma_2$.
In this rule, $\gamma$ is the control point of the program where the function call is made, $\gamma_1$ is the entry point of the called procedure,
and $\gamma_2$ is the return point of the call.
A rule $r_2$ models a return, whereas a rule $r_3$ corresponds to \changeifma{a \textit{simple} statement} (neither a call nor a return). A configuration of $\mathcal{P}$ is a pair $\langle p, \omega \rangle$, where $p$ is a control location and $\omega \in \Gamma^*$ is the stack content.  \changeifma{\changenaad{For technical \changenaap{reasons}, we suppose w.l.o.g.} that the bottom stack symbol $\sharp$ is never popped from the stack,} i.e., there is no rule in the form  $\langle p, \sharp \rangle \changenaad{\xrightarrow{t}} \langle q, \omega\rangle \in \Delta$ ($t \in \{call, ret, int\}$). ${\mathcal{P}} $ defines a transition relation \changenaal{\changeifma{$\xRightarrow{}_{\mathcal{P}}$} $(t \in \{call, ret, int\})$} as follows:
If $\langle p, \gamma \rangle \xrightarrow{t} \langle q, \omega \rangle$, then for every $\omega' \in \Gamma^*$, $\langle p, \gamma \omega' \rangle$ \changenaab{\changeifma{$\xRightarrow{}_{\mathcal{P}}$}} $\langle q, \omega \omega' \rangle$. In other words, $\langle q, \omega \omega' \rangle$ is an immediate successor of $\langle p, \gamma \omega' \rangle$. \changenaas{Let $\xRightarrow{*}_{\mathcal{P}}$ be the reflexive and transitive closure of \changeifma{$\xRightarrow{}_{\mathcal{P}}$}.}

A \changenaad{\changeifma{run}} of $\mathcal{P}$ from $\langle p_0, \omega_0 \rangle$ is \changenaal{a sequence} \changenaab{$\langle p_0, \omega_0 \rangle  \langle p_1, \omega_1 \rangle \langle p_2, \omega_2 \rangle...$ where $\langle p_i, \omega_i \rangle \in P \times \Gamma^*$} \changenaad{s.t.} for every $i \geq 0$, 
\changenaab{$\langle p_i, \omega_i \rangle$ $\changeifma{\xRightarrow{}_{\mathcal{P}}}$ 	$\langle p_{i+1}, \omega_{i+1} \rangle$.} \changenaal{Given a configuration $\langle p, \omega \rangle$, let $\changenaam{Traces(\langle p, \omega \rangle)}$ be the set of all possible \changenaad{\changeifma{runs}} starting from $\langle p, \omega \rangle$.}



\subsection{\changenaad{Global and abstract successors}}
\label{BCARET_sec:globalAbstractCallerPath}
Let $\uppi = \langle p_0, \omega_0 \rangle \langle p_1, \omega_1 \rangle...$ be a \changenaas{\changeifma{run} starting from $\langle p_0, \omega_0 \rangle$}. Over  $\uppi$, \changenaad{two} kinds of successors are defined for every position $\langle p_i, \omega_i \rangle$:

\ifdefined \NotNeedToReducePages \begin{itemize} \else \begin{itemize}[noitemsep,topsep=0pt] \fi 
		\item{\textit{global-successor}: The global-successor of $\langle p_i, \omega_i \rangle$ is $\langle p_{i+1}, \omega_{i+1} \rangle$ \changenaa{where $\langle p_{i+1}, \omega_{i+1} \rangle$ is an immediate successor of $\langle p_i, \omega_i \rangle$}.}
		\item{\textit{abstract-successor}: The abstract-successor of $\langle p_i, \omega_i \rangle$ is determined \changenaab{as follows:}
			\begin{itemize}


				\item{If \changeifma{$\langle p_i, \omega_i \rangle \changeifma{\xRightarrow{}_{\mathcal{P}}} \langle p_{i+1}, \omega_{i+1} \rangle$ corresponds to a call statement, there are two cases: (1) if $\langle p_i, \omega_i \rangle$ has $\langle p_k, \omega_k \rangle$ as a corresponding
						return-point in $\uppi$, then, the abstract successor of $\langle p_i, \omega_i \rangle$ is $\langle p_k, \omega_k \rangle$; (2) if
						$\langle p_i, \omega_i \rangle$ does not have any corresponding return-point in $\uppi$, then, the
						abstract successor of $\langle p_i, \omega_i \rangle$ is $\bot$. }}

				\item{\changenaa{If  \changeifma{$\langle p_i, \omega_i \rangle \changeifma{\xRightarrow{}_{\mathcal{P}}} \langle p_{i+1}, \omega_{i+1} \rangle$ corresponds to a \textit{simple} statement}, the abstract successor of $\langle p_i, \omega_i \rangle$ is $\langle p_{i+1}, \omega_{i+1} \rangle$.}}
				
				\item{\changenaa{If \changeifma{$\langle p_i, \omega_i \rangle \changeifma{\xRightarrow{}_{\mathcal{P}}} \langle p_{i+1}, \omega_{i+1} \rangle$ corresponds to a return statement}, the abstract successor of $\langle p_i, \omega_i \rangle$ is defined as $\bot$.}}

				%
				%

				%
			\end{itemize}
		}
	\end{itemize}

	\begin{figure*}
		\centering
		\scriptsize
		\begin{tikzpicture}[xscale=1, yscale=1]
		\tikzset{lineStyle/.style={blue, thick, ->}};
		\tikzset{myDot/.style={blue, fill = blue, thick}};
		\tikzset{myRectangleNode/.style={rectangle, thick, draw= black, below right, black}};
		\draw [lineStyle] (0, 0) -- (1,0);
		\draw [lineStyle] (1,0) -- (2,0);
		\draw [lineStyle] (2,0) -- (2.5, -1);
		\draw [lineStyle] (2.5, -1) -- (3.5, -1);
		\draw [lineStyle] (3.5, -1) -- (4.5, -1);
		\draw [lineStyle] (4.5, -1) -- (5, -2);
		\draw [lineStyle] (5, -2) -- (6, -2);
		\draw [lineStyle] (6, -2) -- (7, -2);
		\draw [lineStyle] (7, -2) -- (7.5, -1) -- (8.5, -1);
		\draw [dotted] (8.5, -1) -- (9.5, -1);
		\draw [lineStyle] (10.5, -1) -- (11, 0);
		\draw [dotted] (11, 0) -- (12, 0);
		\draw [myDot] (0,0) circle [radius=0.04];
		\draw [myDot] (1,0) circle [radius=0.04];
		\draw [myDot] (2,0) circle [radius=0.04]; 
		\draw [myDot] (2.5, -1) circle [radius=0.04];
		\draw [myDot] (3.5, -1) circle [radius=0.04];
		\draw [myDot] (4.5, -1) circle [radius=0.04];
		\draw [myDot] (5, -2) circle [radius=0.04];
		\draw [myDot] (6, -2) circle [radius=0.04];
		\draw [myDot] (7, -2) circle [radius=0.04]; 
		\draw [myDot] (7.5, -1) circle [radius=0.04];
		\draw [myDot] (8.5, -1) circle [radius=0.04];
		\draw [myDot] (10.5, -1) circle [radius=0.04];
		\draw [myDot] (11, 0) circle [radius=0.04];
		\draw [myDot] (12, 0) circle [radius=0.04];
		\node[below ]() at (0,0) {\scriptsize $ \langle p_0, \omega_0 \rangle$};
		\node[below ]() at (1,0) {$\langle p_1, \omega_1 \rangle$};
		\node[above ]() at (2,0) {$\langle p_2, \omega_2 \rangle$};
		\node[below ]() at (2.5,-1) {$\langle p_3, \omega_3 \rangle$};
		\node[below ]() at (3.5, -1) {$\langle p_4, \omega_4 \rangle$};
		\node[above ]() at (4.5, -1) {$\langle p_5, \omega_5 \rangle$};
		\node[below ]() at (5, -2) {$\langle p_6, \omega_6 \rangle$};
		\node[above ]() at (6, -2) {$\langle p_7, \omega_7 \rangle$};
		\node[below ]() at (7, -2) {$\langle p_8, \omega_8 \rangle$};
		\node[above ]() at (7.5, -1) {$\langle p_9, \omega_9 \rangle$};
		\node[below ]() at (8.5, -1) {$\langle p_{10}, \omega_{10} \rangle$};
		\node[above ]() at (11,0){$\langle p_k, \omega_k \rangle$};

		\node[above]() at (0.5,0) {$\changenaa{int}$};
		
		\node[left]() at (2.25,-.5) {$\changenaa{call}$};

		\node[left]() at (4.75, -1.5) {$\changenaa{call}$};

		\node[right]() at (7.5, -1.5) {$\changenaa{ret}$};

		%
		\node[above left](node1) at (2,0) {};
		\node[above left](node1) at (11,0) {};
		\draw[dotted, red, thick, ->] (2,0).. controls(6,0.5).. (11,0);
		\draw[dotted, red, thick, ->] (4.5,-1).. controls(5.5, -0.8).. (7.5, -1);


		\node[right]() at (1, -1.5) {global-successor};
		\draw[lineStyle] (0, -1.5) -- (1, -1.5);
		\node[right]() at (1, -2) {abstract-successor};
		\draw[->, dotted, red, thick] (0, -2) -- (1, -2);

		\end{tikzpicture}
		\caption{Two kinds of successors on a \changenaad{\changeifma{run}}
		} \label{BCARET_fig:terminology}
	\end{figure*}
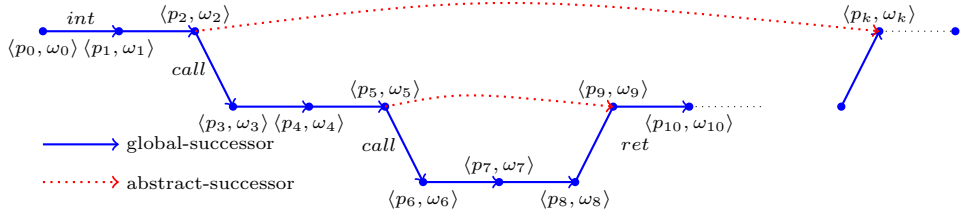
	
	\noindent
	For example, in Figure \ref{BCARET_fig:terminology}:
	\ifdefined \NotNeedToReducePages \begin{itemize} \else \begin{itemize}[noitemsep,topsep=0pt] \fi 

			\item{The \changenaal{global-successors} of $\langle p_1, \omega_1 \rangle$ and $\langle p_2, \omega_2 \rangle$ are $\langle p_2, \omega_2 \rangle$ and $\langle p_3, \omega_3 \rangle$ respectively.}
			\item{The \changenaal{abstract-successors} of $\langle p_2, \omega_2 \rangle$ and $\langle p_5, \omega_5 \rangle$ are $\langle p_k, \omega_k \rangle$ and $\langle p_9, \omega_9 \rangle$ respectively.}

		\end{itemize}

		\medskip
		\noindent
		\changenaad{Let $\langle p, \omega \rangle$ be a configuration of a PDS $\mathcal{P}$.  A configuration $\langle p', \omega' \rangle $ is defined as a global-successor of  $\langle p, \omega \rangle$ iff $ \langle p', \omega' \rangle $ is a global-successor of  $\langle p, \omega \rangle$ over a \changeifma{run} $\uppi \in   \changenaam{Traces(\langle p, \omega \rangle)}$. Similarly, a configuration $\langle p', \omega' \rangle $ is defined as an abstract-successor of  $\langle p, \omega \rangle$ iff $ \langle p', \omega' \rangle $ is an abstract-successor of  $\langle p, \omega \rangle$ over a \changeifma{run} $\uppi \in   \changenaam{Traces(\langle p, \omega \rangle)}$}

		

		
		\medskip
		\noindent
		\changenaad{A \textit{global-path} of $\mathcal{P}$ from $\langle p_0, \omega_0 \rangle$ is \changenaas{a sequence} $\langle p_0, \omega_0 \rangle  \langle p_1, \omega_1 \rangle \langle p_2, \omega_2 \rangle...$ where $\langle p_i, \omega_i \rangle \in P \times \Gamma^*$ s.t. for every $i \geq 0$, $\langle p_{i+1}, \omega_{i+1} \rangle$ is a global-successor of $\langle p_i, \omega_i \rangle$. Similarly, an \textit{abstract-path} of $\mathcal{P}$ from $\langle p_0, \omega_0 \rangle$ is a sequence $\langle p_0, \omega_0 \rangle  \langle p_1, \omega_1 \rangle \allowbreak \langle p_2, \omega_2 \rangle...$ where $\langle p_i, \omega_i \rangle \in P \times \Gamma^*$ s.t. for every $i \geq 0$, $\langle p_{i+1}, \omega_{i+1} \rangle$ is an abstract-successor of $\langle p_i, \omega_i \rangle$.} \changeallfullversionb{For instance, in Figure \ref{BCARET_fig:terminology}, $\langle p_0, \omega_0 \rangle \langle p_1, \omega_1 \rangle \langle p_2, \omega_2 \rangle \allowbreak \langle p_3, \omega_3 \rangle \langle p_4, \omega_4 \rangle \langle p_5, \omega_5 \rangle...$ is a global-path, while $\langle p_0, \omega_0 \rangle \langle p_1, \omega_1 \rangle \langle p_2, \omega_2 \rangle \langle p_k, \omega_k \rangle...$ is an abstract-path.}

		\ifdefined \ShowDraftThing 

		\begin{lstlisting}[basicstyle=\ttfamily\footnotesize, tabsize=2, language=C,xleftmargin=0.5em, escapeinside={(*}{*)}, multicols=1,breaklines=true,caption={A sample program},label={lst:exampleCode}]
		
		
		
		void foo()
		
		(* $f_0:$ *)  fopen("file1", "r");
		(* $f_1:$ *)  if (?)
		(* $f_2:$ *)  	read;
		(* $f_3:$ *)  else
		(* $f_4:$ *)  	foo();
		(* $f_5:$ *) write;
		(* $f_6:$ *) return;
		
		
		
		\end{lstlisting}
		
		\begin{tikzpicture}
		[node distance= 2cm, auto]
		\tikzset {
			node1/.style={rectangle, draw, text centered}
		}
		
		\node at (0,0) (p0) {$(\langle p_0, \omega_0 \rangle, f_0)$};
		
		\node[below of=p0] 		(p1) {$(\langle p_1, \omega_1 \rangle, f_1)$};
		\node[below left of=p1] 		(p2) {$(\langle p_2, \omega_2 \rangle, f_2)$};
		\node[below left of=p2] 		(p3) {$(\langle p_3, \omega_3 \rangle, f_5)$};
		\node[below left of=p3] 		(p4) {$(\langle p_4, \omega_4 \rangle, f_6)$};

		\draw[->](p0)--(p1) node[myrectangle,left=1mm]{$(fopen, int)$};
		\draw[->](p1)--(p2) node[myrectangle,left=1mm]{$(read, int)$};
		\draw[->](p2)--(p3) node[myrectangle,left=1mm]{$(write, int)$};
		\draw[->](p3)--(p4) node[myrectangle,left=1mm]{$(return, ret)$};

		\node[below right of=p1] 		(p5) {$(\langle p_5, \omega_5 \rangle, f_4)$};
		\node[below right of=p5] 		(p6) {$(\langle p_6, \omega_63 \rangle, f_0)$};

		\end{tikzpicture}

		\todo{Intuitively, abstract paths express paths in the same procedure, global paths..}
		
		\todo{Formally define global-path, abstract path, define formally global path from one configuration. Maybe not necessary}
		
		\todo{draw a \changeifma{run} tree with configurations, several runs, don't need to show an C, just a \changeifma{run} tree with call,int,.. and tell intuitively}
		
		\notesctl {I have changed this to define it directly on PDSs. But I still think that we are loosing semantics of CTL when pick up one path and illustrate like this. The point is how to define global successor, abstract successor of a configuration. E.g, Look at Figure 1, it's ok to suppose that $\uppi$ is a path starting from $p_0 \omega_0$, but it is too constraint, loose information to illustratet the global path starting from  $p_0 \omega_0$ by showing a path on $\uppi$ because a global path from $p_2 \omega_2$ may be NOT belongs to $\uppi$. Actually, a global path from $p_0 \omega_0$  should be obtained inductively by the edge from $p_0 \omega_0$  to $p_1 \omega_1$ and select one global path from $p_1 \omega_1$. Then apply the same definition for  $p_1 \omega_1$. However, this is just a small point, we can change this quickly, nothing changes for others Sections. }

		\subsection{Modelling Programs as Labelled Pushdown Systems}

		\begin{lstlisting}[basicstyle=\ttfamily\footnotesize, tabsize=2, language=C,xleftmargin=0.5em, escapeinside={(*}{*)}, multicols=2,breaklines=true,caption={A sample program},label={lst:exampleCode}]
		
		
		
		void foo()
		
		(* $f_0:$ *)  fopen("file1", "r");
		(* $f_1:$ *)  	if (?)
		(* $f_2:$ *)  	 read;
		(* $f_3:$ *)  	else
		(* $f_4:$ *)  		copy;
		(* $f_5:$ *) 	foo2();
		(* $f_6:$ *) 	return;
		
		
		
		
		
		void foo2()
		
		(* $f'_0:$ *)  	fopen("file2");
		(* $f'_1:$ *)  		.....
		(* $f'_2:$ *)  		......
		(* $f'_3:$ *)  	fclose(file2);
		(* $f'_4:$ *)  	return;
		
		
		
		
		
		
		
		
		
		
		\end{lstlisting}

		%
		%
		%
		%
		%
		%
		%
		%
		%
		%
		%
		%
		%
		
		\begin{tikzpicture}[>=stealth]
		
		\node[myellipse](0){0};
		\node[myellipse,below=of 0](1){1};
		\node[myellipse,below left=of 1](2){2};
		\node[myellipse,below=of 2](3){3};
		\node[myellipse,below right=of 3](ret){ret};

		\node[myellipse,below right=of 1](4){4};
		\node[myellipse,below=of 4](5){5};

		\draw[->](0)--(1)node[myrectangle,left=1mm]{$fopen$};
		\draw[->](1)--(2)node[myrectangle,left=1mm]{$read$};
		\draw[->](2)--(3)node[myrectangle,left=1mm]{\textbf{call} $foo2$};
		\draw[->](3)--(ret);

		\draw[->](1)--(4)node[myrectangle,right=1mm]{$write$};
		\draw[->](4)--(5)node[myrectangle,right=1mm]{\textbf{call} $foo2$};
		\draw[->](5)--(ret);
		

		\end{tikzpicture}

		\todo{Maybe we should show a translation from a simple program to a Tree here to illustrate how it is useful to define BCARET}

		\fi

		
		\subsection{Multi Automata}
		
		\begin{definition}
			\cite{DBLP:conf/concur/BouajjaniEM97} Let \changeallfullversionb{$\mathcal{P} = (P, \Gamma, \Delta, \sharp)$} \changenaas{be a PDS}. A $\mathcal{P} $-Multi-Automaton (MA for short) is a tuple $\mathcal{A} = (Q, \Gamma, \delta, I, Q_f)$, where $Q$ is a finite set of states, $\delta \subseteq Q \times \Gamma \times Q$ is a finite set of transition rules, \changenaas{$I = P \subseteq Q$} is a set of initial states, $Q_f \subseteq Q$ is a set of final states.
		\end{definition}	
		
		The transition relation $\changenaae{\xrightarrow{}_\delta} \subseteq Q \times \Gamma^* \times Q$ is defined as \changenaac{follows}:
		
		\ifdefined \NotNeedToReducePages \begin{itemize} \else \begin{itemize}[noitemsep,topsep=0pt] \fi 
				\item {$q \changenaae{\xrightarrow{\epsilon}_\delta} q$ for every $q \in Q$ }
				\item {$q \changenaae{\xrightarrow{\gamma}_\delta} q'$ if $(q, \gamma, q') \in \delta$ }
				\item {if $q \changenaae{\xrightarrow{\omega}_\delta} q'$ and $q' \changenaae{\xrightarrow{\gamma}_\delta} q''$, then, $q \changenaae{\xrightarrow{\omega\gamma}_\delta} q''$}
			\end{itemize}
			
			$\mathcal{A}$ recognizes a configuration $\langle p, \omega \rangle$ \changenaas{where $p \in P$, $\omega \in \Gamma^*$} iff $\changenaas{p \xrightarrow{\omega}_\delta} q$ for some $q \in Q_f$. \changenaar{The language of $\mathcal{A}$, $L (\mathcal{A})$, is the set of all configurations which are recognized by $\mathcal{A}$}. A set of configurations is \textit{regular} if it is recognized by some \changenaas{Multi-Automaton}.
			

			\section{Branching Temporal Logic of Calls and Returns - BCARET}
			\label{BCARET_sec:BCARETDefinition}

			In this section, we define \changenaas{the Branching temporal logic of CAlls and RETurns BCARET}. For technical reasons, we assume w.l.o.g. that \changenaab{BCARET} formulas are given in positive normal form, i.e. negations are applied only to atomic propositions. To do that, we use \changenaal{the} \changenaad{release} operator \changenaap{$R$} as a dual of the until operator \changenaap{$U$}. 
			
			\begin{definition}{\textbf{Syntax of BCARET}}
				
				\noindent
				\changenaad{Let $AP$ be a finite set of atomic propositions, a BCARET} formula $\varphi$ is defined as follows, \changenaas{where $b \in \{g, a\}$, $e \in AP$:}
				
				$$\varphi::=  \changenaad{true \;|\; false \;|\; } e \;|\;  \neg e \;|\; \varphi \vee \varphi \;|\; \varphi \wedge
				\varphi \;|\; EX^b \varphi \;|\; A X^b \varphi \;|\; E [\varphi U^b \varphi] \;|\; A [\varphi U^b \varphi] \;|\; E [\varphi R^b \varphi] \;|\; A [\varphi R^b \varphi ] $$

			\end{definition}

			%

			\changenaad{Let $\mathcal{P} = (P, \Gamma, \Delta, \changenaa{\sharp})$ be a PDS, $\lambda: AP \rightarrow 2^{P \times \Gamma^*}$ be a labelling function that assigns \changenaal{to} each atomic proposition $e \in AP$ a set of configurations of $\mathcal{P}$. The satisfiability relation of a BCARET formula $\varphi$ at a configuration $\changenaap{\langle p_0, \omega_0 \rangle}$ w.r.t. the labelling function $\lambda$, denoted by $\changenaap{\langle p_0, \omega_0 \rangle} \vDash_{\lambda} \varphi$, is defined inductively as follows:}
			
			\ifdefined \NotNeedToReducePages \begin{itemize} \else \begin{itemize}[noitemsep,topsep=0pt] \fi 
					\item {\changenaad{$\changenaap{\langle p_0, \omega_0 \rangle} \vDash_{\lambda} true$ for every $\changenaap{\langle p_0, \omega_0 \rangle} $}}
					\item {\changenaad{$\changenaap{\langle p_0, \omega_0 \rangle} \nvDash_{\lambda} false$ for every $\changenaap{\langle p_0, \omega_0 \rangle} $}}
					
					\item {$\changenaap{\langle p_0, \omega_0 \rangle} \vDash_{\lambda} e \changenaad{\;(e \in AP)}$ iff $\changenaap{\langle p_0, \omega_0 \rangle}  \in \lambda(e)$}
					\item {$\changenaap{\langle p_0, \omega_0 \rangle} \vDash_{\lambda} \neg e \changenaad{\;(e \in AP)} $ iff $\changenaap{\langle p_0, \omega_0 \rangle}  \notin \lambda(e)$}
					\item {$\changenaap{\langle p_0, \omega_0 \rangle} \vDash_{\lambda} \varphi_1 \vee \varphi_2 $ iff ($\changenaap{\langle p_0, \omega_0 \rangle} \vDash_{\lambda} \varphi_1$ or $\changenaap{\langle p_0, \omega_0 \rangle} \vDash_{\lambda} \varphi_2$)}
					\item {$\changenaap{\langle p_0, \omega_0 \rangle} \vDash_{\lambda} \varphi_1 \wedge \varphi_2 $ iff ($\changenaap{\langle p_0, \omega_0 \rangle} \vDash_{\lambda} \varphi_1$ and $\changenaap{\langle p_0, \omega_0 \rangle} \vDash_{\lambda} \varphi_2$)}

					\item {$\changenaap{\langle p_0, \omega_0 \rangle} \vDash_{\lambda} E X^g \varphi $ iff there exists a global-successor $\langle p', \omega' \rangle$ of $\changenaap{\langle p_0, \omega_0 \rangle}$ such that $\changenaad{ \langle p', \omega' \rangle } \vDash_{\lambda} \varphi $  }

					\item {$\changenaap{\langle p_0, \omega_0 \rangle} \vDash_{\lambda} A X^g \varphi $ iff  $\changenaad{ \langle p', \omega' \rangle } \vDash_{\lambda} \varphi $ for every global-successor $\langle p', \omega' \rangle$ of $\changenaap{\langle p_0, \omega_0 \rangle}$ }

					\item {$\changenaap{\langle p_0, \omega_0 \rangle} \vDash_{\lambda} E[\varphi_1 U^g \varphi_2] $ iff there exists a global-path \changenaal{$\uppi$ = $\langle p_0, \omega_0 \rangle \langle p_1, \omega_1 \rangle \langle p_2, \omega_2 \rangle...$ of $\mathcal{P}$ starting from $\changenaap{\langle p_0, \omega_0 \rangle}$ \changenaap{s.t.}} $\exists i \geq 0$, $\langle p_i, \omega_i \rangle \vDash_{\lambda} \varphi_2$ and for every $0 \leq j < i$,  $\langle p_j, \omega_j \rangle \vDash_{\lambda} \varphi_1$  }

					\item {$\changenaap{\langle p_0, \omega_0 \rangle} \vDash_{\lambda} A[\varphi_1 U^g \varphi_2] $ iff for every global-path \changenaal{ $\uppi$ = $\langle p_0, \omega_0 \rangle \langle p_1, \omega_1 \rangle \langle p_2, \omega_2 \rangle...$of $\mathcal{P}$ starting from $\changenaas{\langle p_0, \omega_0 \rangle}$\changenaas{,}} $\exists i \geq 0$, $\langle p_i, \omega_i \rangle \vDash_{\lambda} \varphi_2$ and for every $0 \leq j < i$,  $\langle p_j, \omega_j \rangle \vDash_{\lambda} \varphi_1$  }

					\item {$\changenaap{\langle p_0, \omega_0 \rangle} \vDash_{\lambda} E[\varphi_1 R^g \varphi_2] $ iff there exists a global-path \changenaal{$\uppi$ = $\langle p_0, \omega_0 \rangle \langle p_1, \omega_1 \rangle \langle p_2, \omega_2 \rangle...$ of $\mathcal{P}$ starting from $\changenaap{\langle p_0, \omega_0 \rangle}$  \changenaap{s.t.}} \changenaal{ for every $i \geq 0$,  if $\langle p_i, \omega_i \rangle \nvDash_\lambda \varphi_2$ then there exists  $0 \leq j <i $ s.t.  $\langle p_j, \omega_j \rangle \vDash_{\lambda} \varphi_1$} }

					\item {$\changenaap{\langle p_0, \omega_0 \rangle} \vDash_{\lambda} A[\varphi_1 R^g \varphi_2] $ iff for every global-path \changenaal{ $\uppi$ = $\langle p_0, \omega_0 \rangle \langle p_1, \omega_1 \rangle \langle p_2, \omega_2 \rangle...$ of $\mathcal{P}$ starting from $\changenaas{\langle p_0, \omega_0 \rangle}$\changenaas{,}} \changenaal{ for every $i \geq 0$,  if $\langle p_i, \omega_i \rangle \nvDash_\lambda \varphi_2$ then there exists  $0 \leq j <i $ s.t.  $\langle p_j, \omega_j \rangle \vDash_{\lambda} \varphi_1$} }

					
					\item {$\changenaap{\langle p_0, \omega_0 \rangle} \vDash_{\lambda} E X^a \varphi $ iff there exists an abstract-successor $\langle p', \omega' \rangle$ of $\changenaap{\langle p_0, \omega_0 \rangle}$ such that $\langle p', \omega' \rangle \vDash_{\lambda} \varphi $  }
					
					\item {$\changenaap{\langle p_0, \omega_0 \rangle} \vDash_{\lambda} A X^a \varphi $ iff  $\changenaad{ \langle p', \omega' \rangle } \vDash_{\lambda} \varphi $ for every abstract-successor $\langle p', \omega' \rangle$ of $\changenaap{\langle p_0, \omega_0 \rangle}$ }

					\item {$\changenaap{\langle p_0, \omega_0 \rangle} \vDash_{\lambda} E[\varphi_1 U^a \varphi_2] $ iff there exists an abstract-path \changenaal{$\uppi$ = $\langle p_0, \omega_0 \rangle \langle p_1, \omega_1 \rangle \langle p_2, \omega_2 \rangle...$ of $\mathcal{P}$ starting from $\changenaap{\langle p_0, \omega_0 \rangle}$ \changenaap{s.t.}} $\exists i \geq 0$, $\langle p_i, \omega_i \rangle \vDash_{\lambda} \varphi_2$ and for every $0 \leq j < i$,  $\langle p_j, \omega_j \rangle \vDash_{\lambda} \varphi_1$}

					\item {$\changenaap{\langle p_0, \omega_0 \rangle} \vDash_{\lambda} A[\varphi_1 U^a \varphi_2] $ iff for every abstract-path \changenaas{$\uppi$ = $\langle p_0, \omega_0 \rangle \langle p_1, \omega_1 \rangle \langle p_2, \omega_2 \rangle...$ of $\mathcal{P}$\changenaas{,} $\exists i \geq 0$, $\langle p_i, \omega_i \rangle \vDash_{\lambda} \varphi_2$ and for every $0 \leq j < i$,  $\langle p_j, \omega_j \rangle \vDash_{\lambda} \varphi_1$  }}

					\item {$\changenaap{\langle p_0, \omega_0 \rangle} \vDash_{\lambda} E[\varphi_1 R^a \varphi_2] $ iff there exists an abstract-path \changenaal{$\uppi$ = $\langle p_0, \omega_0 \rangle \langle p_1, \omega_1 \rangle \langle p_2, \omega_2 \rangle...$ of $\mathcal{P}$ starting from $\changenaap{\langle p_0, \omega_0 \rangle}$ \changenaap{s.t.}} \changenaal{ for every $i \geq 0$,  if $\langle p_i, \omega_i \rangle \nvDash_\lambda \varphi_2$ then there exists  $0 \leq j <i $ s.t.  $\langle p_j, \omega_j \rangle \vDash_{\lambda} \varphi_1$}   }
					
					\item {$\changenaap{\langle p_0, \omega_0 \rangle} \vDash_{\lambda} A[\varphi_1 R^a \varphi_2] $ iff for every abstract-path \changenaal{$\uppi$ = $\langle p_0, \omega_0 \rangle \langle p_1, \omega_1 \rangle \langle p_2, \omega_2 \rangle...$  of $\mathcal{P}$ starting from $\changenaas{\langle p_0, \omega_0 \rangle}$\changenaas{,}} \changenaal{ for every $i \geq 0$,  if $\langle p_i, \omega_i \rangle \nvDash_\lambda \varphi_2$ then there exists  $0 \leq j <i $ s.t.  $\langle p_j, \omega_j \rangle \vDash_{\lambda} \varphi_1$}}
					
				\end{itemize}

				\notesctl{... means that there are more formulas can be rewritten in the same way, do I need to write all?}

				\changenaaq{Other BCARET operators can be expressed by the above operators: $EF^g \varphi = E[true \; U^g \varphi]$, $EF^a \varphi = E[true \; U^a \varphi]$, $AF^g \varphi = A[true \; U^g \varphi]$, $AF^a \varphi = A[true U^a \varphi]$,...}

				\medskip
				\noindent
				\textbf{Closure.} \changenaas{Given a BCARET formula $\varphi$,  the closure $Cl(\varphi)$ is the set of all subformulae of $\varphi$, including $\varphi$.} 
				
				\medskip
				\noindent
				\textbf{\changenaat{Regular Valuations.}} \changenaas{We talk about \changenaat{regular valuations} when for every $e \in AP$, $\lambda(e)$ is a regular language.}

				\begin{remark}
					\changenaas{ CTL can be seen as the subclass of BCARET where the operators $EX^a \varphi, A X^a \varphi, E [\varphi U^a \varphi], A [\varphi U^a \varphi], E [\varphi R^a \varphi], A [\varphi R^a \varphi ]$ are not considered.}
				\end{remark}

				\section{Application}
				\label{BCARET_sec:malwarespecif}
				In this section, we show how BCARET can be used to describe branching-time malicious behaviors.


				\medskip
				\noindent
				\textbf{Spyware Behavior.} The typical behaviour of a spyware is hunting for personal information (emails, bank account information,...) on local drives by searching files matching certain conditions. To do that, it has to search directories of the host to look for interesting files whose names match a specific condition. When a file is found, the spyware will invoke a payload to steal the information, then continue looking for the remaining matching files. When a folder is found, it will enter the folder path and continue scanning that folder recursively. To achieve this behavior, the spyware first calls the API function $FindFirstFileA$ to search for the first matching file in a given folder path. After that, it has to check whether the call to the API function $FindFirstFileA$ succeeds or not. If the function call fails, the spyware will call the function $GetLastError$. Otherwise, if the function call is successful, $FindFirstFileA$ will return a search handle $h$.  There are two possibilities in this case. If the returned result is a folder, it will call the API function $FindFirstFileA$ again to search for matching results in the found folder. If the returned result is a file, it will call the API function $FindNextFileA$ using $h$ as first parameter to look for the remaining matching files. This behavior cannot be expressed by LTL or CTL because it requires to express that the return value of the function $FindFirstFileA$ should be used as input to the API function $FindNextFileA$. It cannot be described by CARET neither (because this is a branching-time property). Using BCARET, the above behavior can be expressed by the following formula:

				\begin{align*}
				\varphi_{sb} = \bigvee_{d \in D}  EF^g  & \Bigg(call(FindFirstFileA) \wedge EX^a (eax = d) \wedge AF^a \\
				& \quad \bigg(call(GetLastError)  \vee call(FindFirstFileA) \\
				& \qquad \vee \Big(call(FindNextFileA) \wedge d \Gamma^* \Big) \bigg) \Bigg)
				\end{align*}

				%
				%
				%
				%

				\noindent
				where the $\bigvee$ is taken over all possible memory addresses $d$ which
				contain the values of search handles $h$ in the program, $EX^a$ is a BCARET operator that means "next in some run, in the same procedural context"; $EF^g$ is the standard CTL $EF$ operator (eventually in some run), while $AF^a$ is a BCARET operator that means "eventually in all runs, in the same procedural context".

				
				\changeifma{In binary codes and assembly programs}, the return value of an API function is put in the register $eax$. Thus, the return value of $FindFirstFileA$ is the value of $eax$ at its corresponding return-point. Then, the subformula $ (\text{call(FindFirstFileA)} \wedge EX^{a} (eax=d))$ states that there is a call to the API $FindFirstFileA$ and the return value of this function is $d$ (the abstract successor of a call is its corresponding return-point). When \textit{FindNextFileA} is invoked, it requires a search handle as parameter and this search handle must be put on top of the program stack (since parameters are passed through the stack in assembly). The requirement that $d$ is on top of the program stack is expressed by the regular expression $d \Gamma^*$. Thus, the subformula $[\text{call(FindNextFileA)} \wedge d \Gamma^*]$ expresses that \textit{FindNextFileA} is called with $d$ as parameter ($d$ stores the information of the search handle). Therefore, $\varphi_{sb}$ expresses then that there is a call to the API $FindFirstFileA$ with the return value $d$ (the search handle), then, in all runs starting from that call, there will be either a call to the API function $GetLastError$ or a call to the function $FindFirstFileA$ or a call to the function $FindNextFileA$ in which $d$ is used as a parameter.


				\changeifma{To detect spyware, \cite{DBLP:conf/SAC/VuTayssir2017}  used the following CARET formula:}
				
				\noindent
				\resizebox{\hsize}{!}{ $\varphi'_{sb} = \bigvee_{d\in D} F^{g} (\text{call(FindFirstFileA)} \wedge X^{a} (eax = d) \wedge F^{a} (\text{call(FindNextFileA)} \wedge d \Gamma^*))$ }

				\medskip
				\noindent
				\changeifma{It can be seen that this CARET formula $\varphi'_{sb}$ is not as precise as the BCARET formula $\varphi_{sb}$,  as it does not deal with the case when the returned result of $FindFirstFileA$ is a folder or an error. Thus, this CARET formula $\varphi'_{sb}$ may lead to false alarms that can be avoided using our BCARET formula $\varphi_{sb}$.  BCARET can deal with it because BCARET is a branching-time temporal logic. For example, $AF^a$ allows us to take into account all possible abstract-paths from a certain state in the \changeifma{computation} tree. By using  $AF^a$, $\varphi_{sb}$ can deal with different returned values of $FindFirstFileA$ as presented above.}


				\section{BCARET Model-Checking for Pushdown Systems}
				
				In this section, we consider "standard" BCARET model-checking for pushdown systems where an atomic proposition holds at a configuration $c$ or not \changenaab{depends} only on the control state of $c$, not on its stack.

				
				\subsection{Alternating B\"{u}chi Pushdown Systems (ABPDSs).}
				
				\begin{definition}
					An Alternating B\"{u}chi Pushdown System (ABPDS) is a tuple $\mathcal{BP} = (P, \Gamma, \Delta, F)$, where $P$ is a set of control locations, $\Gamma$ is \changenaas{the stack alphabet}, $F \subseteq P$ is a set of accepting control locations \changenaal{and}  $\Delta$ is a transition function that maps each element of $P \times \Gamma$ with a positive boolean formula over $P \times \Gamma^*$.
				\end{definition}


				
				A configuration of $\mathcal{BP}$ is a pair $\langle p, \omega \rangle$, where $p \in P$ is the current control location and $\omega \in \Gamma^*$ is the current stack content. Without loss of generality, we suppose that the boolean formulas of ABPDSs are in disjunctive normal form $\bigvee_{j=1}^{n} \bigwedge_{i=1}^{m_j} \langle p_i^j, \omega_i^j \rangle$. Then, we can \changenaap{see} $\Delta$ as a subset of $(P \times \Gamma) \times 2^{P \times \Gamma^*}$ by rewriting the rules of $\Delta$ in the form $\langle p, \gamma \rangle \rightarrow \bigvee_{j=1}^{n} \bigwedge_{i=1}^{m_j} \langle p_i^j, \omega_i^j \rangle$ as $n$ rules of the form  $\langle p, \gamma \rangle \rightarrow \{\langle p_1^j, \omega_1^j \rangle,..., \langle p_{m_j}^j, \omega_{m_j}^j \rangle\} $, where $1 \leq j \leq n$. Let $\langle p, \gamma \rangle \rightarrow \{\langle p_1, \omega_1 \rangle,..., \langle p_n, \omega_n \rangle \}$ be a rule of $\Delta$, then, for every $\omega \in \Gamma^*$, the configuration $\langle p, \gamma \omega \rangle$\changenaag{(resp.  $\{\langle p_1, \omega_1 \omega \rangle,..., \langle p_n, \omega_n \omega \rangle \}$)} is an immediate predecessor \changenaag{(resp. successor)} of $\{\langle p_1, \omega_1 \omega \rangle,..., \langle p_n, \omega_n \omega \rangle \}$ \changenaag{(resp. $\langle p, \gamma \omega \rangle$)}. 
				
				A run $\rho$ of $\mathcal{BP}$ starting form an initial configuration $\langle p_0, \omega_0 \rangle$ is a tree \changenaas{whose root is} labelled by $\langle p_0, \omega_0 \rangle$, and \changenaal{whose other nodes} are labelled by elements in $P \times \Gamma^*$. If a node of $\rho$ is labelled by a configuration $\langle p, \omega \rangle$ and has $n$ children labelled by \changenaag{$\langle p_1, \omega_1 \rangle,..., \langle p_n, \omega_n \rangle$} respectively, then, $\langle p, \omega \rangle$ must be a predecessor of $\{ \langle p_1, \omega_1 \rangle,..., \langle p_n, \omega_n \rangle\}$ in $\mathcal{BP}$. A path of a run $\rho$ is an infinite sequence of configurations $c_0 c_1 c_2...$ s.t. $c_0$ is the root of $\rho$ and $c_{i+1}$ is one of the children of $c_i$ for every $i \geq 0$. A path is accepting iff it visits infinitely often configurations with control locations in $F$. A run $\rho$ is accepting iff every \changenaal{path of $\rho$ is accepting. The language of} $\mathcal{BP}$, $\mathcal{L}(\mathcal{BP})$, is the set of configurations $c$ s.t. $\mathcal{BP}$ has \changenaap{an accepting run} starting from $c$.

				
				\changenaap{$\mathcal{BP}$ defines the reachability relation $\xRightarrow{}_{\mathcal{BP}_\varphi}$ as follows: (1) $c \xRightarrow{}_{\mathcal{BP}} \{ c\}$ for every $c \in P \times \Gamma^*$, (2) $c \xRightarrow{}_{\mathcal{BP}} C$ if $C$ is an immediate successor of $c$; (3) if $c \xRightarrow{}_{\mathcal{BP}} \{c_1, c_2,..., c_n\}$ and $c_i \xRightarrow{}_{\mathcal{BP}} C_i$ for every $1 \leq i \leq n$, then $c \xRightarrow{}_{\mathcal{BP}} \bigcup_{i=1}^n C_i$.  Given $c_0 \xRightarrow{}_{\mathcal{BP}} C'$, then, $\mathcal{BP}$ has an accepting run from $c_0$ iff $\mathcal{BP}$ has an accepting run from $c'$  for every $c' \in C'$.}
				

				%
				%
				
				\begin{theorem}
					\label{BCARET_theorem:ABPDSMembershipProblem}
					\cite{DBLP:conf/concur/SongT11} \changenaas{Given an ABPDS $\mathcal{BP} = (P, \Gamma, \Delta, F)$, for every configuration $\langle p, \omega \rangle \in P \times \Gamma^*$, whether or not $\langle p, \omega \rangle \in \mathcal{L} (\mathcal{BP})$ can be decided in time $\mathcal{O}(|P|^2.  |\Gamma|. (|\Delta|  2^{5 |P|} + 2^{|P|} |\omega|))$.}

				\end{theorem}

				\subsection{From BCARET model checking of PDSs to the membership problem in ABPDSs}
				\label{BCARET_sec:BCARET_2_Membership}
				Let $\mathcal{P} = (P, \Gamma, \Delta, \sharp)$ be a pushdown system with an initial configuration $c_0$. Given a set of atomic propositions $AP$, let $\varphi$ be a BCARET formula. Let $f: AP \rightarrow 2^P$ be a function that associates each atomic proposition with a set of control states, and $\lambda_f: AP \rightarrow 2^{P \times \Gamma^*}$ be a labelling function s.t. for every $e \in AP$, $\lambda_f(e) = \{\langle p, \omega \changenaad{\rangle} \;|\; p \in f(e), \omega \in \Gamma^* \}$. In this section, we propose an algorithm to check whether $\changeallfullversiona{c_0} \vDash_{\lambda_f} \varphi$. Intuitively, we construct \changenaab{an} Alternating B\"{u}chi Pushdown System \changenaad{$\mathcal{BP}_\varphi$} which \changenaab{recognizes} a configuration $c$ iff  $\changeallfullversiona{c} \vDash_{\lambda_f} \varphi$. Then to check whether $\changeallfullversiona{c_0} \vDash_{\lambda_f} \varphi$, we will check if \changeifma{$c_0 \in \mathcal{L}(\mathcal{BP}_\varphi)$}. The membership problem \changenaad{of an ABPDS} \changenaal{can be} solved effectively by Theorem \ref{BCARET_theorem:ABPDSMembershipProblem}.

				Let $\mathcal{BP}_\varphi = (P', \Gamma', \Delta', F)$ be the ABPDS defined as follows:
				
				\begin{itemize}
					\item {$P' = \changeallfullversionb{ P \cup (P \times Cl(\varphi)) \cup \{ p_{\bot} \}} $}
					\item {$\Gamma' = \changeallfullversionb{\Gamma \cup (\Gamma \times Cl(\varphi)) \cup \{ \gamma_{\bot} \} }$}
					\item {$F = F_1 \cup F_2 \cup F_3$ where 
						
						\begin{itemize}
							\item {$F_1 = \{ \llparenthesis p, e \rrparenthesis \;|\; e \in Cl(\varphi), e \in AP \text{ and } p \in f(e) \} $}
							\item {$F_2 =   \{ \llparenthesis p, \neg e \rrparenthesis \;|\; \neg e \in Cl(\varphi), e \in AP \text{ and } p \notin f(e) \}$}
							\item {$F_3 = \{P \times Cl_{R} (\varphi) \} $ where $Cl_{R} (\varphi)$ is the set of formulas of $Cl(\varphi)$ \changenaal{in the form} $E [\varphi_1 R^b \varphi_2]$ or $A [\varphi_1 R^b \varphi_2]$ \changenaab{($b \in \{g, a\}$)} }
						\end{itemize}
					}
				\end{itemize} 
				
				The transition relation $\Delta'$ is the smallest set of transition rules defined as follows: \changenaas{$\Delta \subseteq \Delta'$ and} for every $p \in P$, $\phi \in Cl(\varphi)$, $\gamma \in \Gamma$, \changenaab{$b \in \{g, a\}$} and \changenaab{$t \in \{call, ret, int\}$}:
				
				\begin{enumerate}[label=($\alpha$\arabic*)]
					\item {If $\phi = e$, $e \in AP$ and $p \in f(e)$, then, $\langle \llparenthesis p, \phi \rrparenthesis, \gamma \rangle \rightarrow \langle \llparenthesis p, \phi \rrparenthesis, \gamma \rangle \in \Delta'$} \label{BCARET_item:standRulesAP}
					
					\item {If $\phi = \neg e$, $e \in AP$ and $p \notin f(e)$, then, $\langle \llparenthesis p, \phi \rrparenthesis, \gamma \rangle \rightarrow \langle \llparenthesis p, \phi \rrparenthesis, \gamma \rangle \in \Delta'$} \label{BCARET_item:standRulesNegAP}
					
					\item {If $\phi = \phi_1 \wedge \phi_2$, then, $\langle \llparenthesis p, \phi \rrparenthesis, \gamma \rangle \rightarrow \langle \llparenthesis p, \phi_1  \rrparenthesis, \gamma \rangle \wedge \langle \llparenthesis p, \phi_2  \rrparenthesis, \gamma \rangle \in \Delta'$} \label{BCARET_item:standRulesAnd}
					
					\item {If $\phi = \phi_1 \vee \phi_2$, then, $\langle \llparenthesis p, \phi \rrparenthesis, \gamma \rangle \rightarrow \langle \llparenthesis p, \phi_1  \rrparenthesis, \gamma \rangle \vee \langle \llparenthesis p, \phi_2  \rrparenthesis, \gamma \rangle \in \Delta'$} \label{BCARET_item:standRulesOr}

					\item {\label{BCARET_item:standRulesEXG} If $\phi = E X^g \phi_1$, then $\llparenthesis p, \phi \rrparenthesis, \gamma \rangle \rightarrow \bigvee_{\langle p, \gamma \rangle \xrightarrow{t} \langle q, \omega \rangle \in \Delta} \langle \llparenthesis q, \phi_1  \rrparenthesis, \omega \rangle  \in \Delta'$ where $t \in \{call, int, ret\}$}
					
					\item {If $\phi = A X^g \phi_1$, then, $\langle \llparenthesis p, \phi \rrparenthesis, \gamma \rangle \rightarrow \bigwedge_{\langle p, \gamma \rangle \xrightarrow{t} \langle q, \omega \rangle \in \Delta} \langle \llparenthesis q, \phi_1  \rrparenthesis, \omega \rangle  \in \Delta'$}\label{BCARET_item:standRulesAXG}
					
					\item {\label{BCARET_item:standRulesEXA} If $\phi = E X^a \phi_1$, then, 	
						\changenaab{$\langle \llparenthesis p, \phi \rrparenthesis, \gamma \rangle \rightarrow  \changenaat{h_1} \vee  \changenaat{h_2} \changeallfullversiona{\vee h_3} \in \Delta'$, where}
						\begin{itemize}
							\item {$\changenaap{\changenaat{h_1} =} \bigvee_{\langle p, \gamma \rangle \xrightarrow{call} \langle q, \gamma_1 \gamma_2 \rangle \in \Delta}  \langle \changenaad{q},  \gamma_1 \llparenthesis \gamma_2, \phi_1 \rrparenthesis \rangle$}
							\item {$\changenaap{\changenaat{h_2} =} \bigvee_{\langle p, \gamma \rangle \xrightarrow{int} \langle q, \omega\rangle   \in \Delta}  \langle \llparenthesis q, \phi_1  \rrparenthesis, \omega \rangle $}
							
							\item {\changeallfullversiona{$\changenaap{\changenaat{h_3} =} \bigvee_{\langle p, \gamma \rangle \xrightarrow{ret} \langle q, \epsilon \rangle   \in \Delta}  \langle \changeallfullversionb{p_{\bot}}, \changeallfullversionb{\gamma_{\bot}} \rangle  $}}
						\end{itemize}
						
					}

					\item {\label{BCARET_item:standRulesAXA} If $\phi = A X^a \phi_1$, then, \changenaad{$\langle \llparenthesis p, \phi \rrparenthesis, \gamma \rangle \rightarrow  \changenaat{h_1} \wedge  \changenaat{h_2} \changeallfullversiona{\wedge h_3} \in \Delta'$, where}
						
						\begin{itemize}
							\item {$\changenaap{\changenaat{h_1} =} \bigwedge_{\langle p, \gamma \rangle \xrightarrow{call} \langle q, \gamma_1 \gamma_2 \rangle \in \Delta}  \langle \changenaad{q},  \gamma_1 \llparenthesis \gamma_2, \phi_1 \rrparenthesis \rangle$}
							\item {$\changenaap{\changenaat{h_2} =} \bigwedge_{\langle p, \gamma \rangle \xrightarrow{int} \langle q, \omega\rangle   \in \Delta}  \langle \llparenthesis q, \phi_1  \rrparenthesis, \omega \rangle $}
							\item {\changeallfullversiona{$\changenaap{\changenaat{h_3} =} \bigwedge_{\langle p, \gamma \rangle \xrightarrow{ret} \langle q, \epsilon \rangle   \in \Delta}  \langle \changeallfullversionb{p_{\bot}}, \changeallfullversionb{\gamma_{\bot}} \rangle  $}}
							
						\end{itemize}
						
					}

					\item { \label{BCARET_item:standRulesEUG} If $\phi = \changenaaf{E [\phi_1 U^g \phi_2]}$, then, 
						
						$\langle \llparenthesis p, \phi \rrparenthesis, \gamma \rangle \rightarrow \langle \llparenthesis p, \phi_2  \rrparenthesis, \gamma \rangle \vee \bigvee_{\langle p, \gamma \rangle \xrightarrow{t} \langle q, \omega \rangle \in \Delta} (\langle \changenaag{\llparenthesis p, \phi_1 \rrparenthesis, \gamma} \rangle \wedge \langle \llparenthesis q, \phi \rrparenthesis, \omega \rangle) \in \Delta'$}

					\item {\label{BCARET_item:standRulesEUA} If $\phi = \changenaaf{E [\phi_1 U^a \phi_2]}$, then, $\langle \llparenthesis p, \phi \rrparenthesis, \gamma \rangle \rightarrow \langle \llparenthesis p, \phi_2  \rrparenthesis, \gamma \rangle \vee \changenaat{h_1} \vee  \changenaat{h_2} \changeallfullversiona{\vee h_3} \in \Delta'$, \changenaas{where}

						\begin{itemize}

							\item {$\changenaap{\changenaat{h_1} =} \bigvee_{\langle p, \gamma \rangle \xrightarrow{call} \langle q, \gamma_1 \gamma_2 \rangle \in \Delta}  \langle \changenaag{\llparenthesis p, \phi_1 \rrparenthesis, \gamma} \rangle \wedge   \langle \changenaad{q},  \gamma_1 \llparenthesis \gamma_2, \phi \rrparenthesis \rangle$}
							\item {$\changenaap{\changenaat{h_2} =} \bigvee_{\langle p, \gamma \rangle \xrightarrow{int} \langle q, \omega\rangle   \in \Delta}  \langle \changenaag{\llparenthesis p, \phi_1 \rrparenthesis, \gamma} \rangle \wedge \langle \llparenthesis q, \phi  \rrparenthesis, \omega \rangle $}
							
							\item {\changeallfullversiona{$\changenaap{\changenaat{h_3} =} \bigvee_{\langle p, \gamma \rangle \xrightarrow{ret} \langle q, \epsilon \rangle   \in \Delta}  \langle \changeallfullversionb{p_{\bot}}, \changeallfullversionb{\gamma_{\bot}} \rangle  $}}
							
						\end{itemize}

					}

					\item {\label{BCARET_item:standRulesAUG} If $\phi = \changenaaf{A [\phi_1 U^g \phi_2]}$, then, 
						
						$\langle \llparenthesis p, \phi \rrparenthesis, \gamma \rangle \rightarrow \langle \llparenthesis p, \phi_2  \rrparenthesis, \gamma \rangle \vee \bigwedge_{\langle p, \gamma \rangle \xrightarrow{t} \langle q, \omega \rangle \in \Delta} (\langle \llparenthesis p, \phi_1  \rrparenthesis, \gamma \rangle \wedge \langle \llparenthesis q, \phi \rrparenthesis, \omega \rangle)\in \Delta'$}

					\item {\label{BCARET_item:standRulesAUA} If $\phi = \changenaaf{A [\phi_1 U^a \phi_2]}$, then, $\langle \llparenthesis p, \phi \rrparenthesis, \gamma \rangle \rightarrow \langle \llparenthesis p, \phi_2  \rrparenthesis, \gamma \rangle \vee \changeallfullversiona{(\changenaat{h_1} \wedge  \changenaat{h_2} \wedge h_3)} \in \Delta'$, \changenaas{where}

						\begin{itemize}

							\item {$\changenaap{\changenaat{h_1} =} \bigwedge_{\langle p, \gamma \rangle \xrightarrow{call} \langle q, \gamma_1 \gamma_2 \rangle \in \Delta}  \langle \changenaag{\llparenthesis p, \phi_1 \rrparenthesis, \gamma} \rangle \wedge   \langle \changenaad{q},  \gamma_1 \llparenthesis \gamma_2, \phi \rrparenthesis \rangle$}
							\item {$\changenaap{\changenaat{h_2} =} \bigwedge_{\langle p, \gamma \rangle \xrightarrow{int} \langle q, \omega\rangle   \in \Delta}  \langle \changenaag{\llparenthesis p, \phi_1 \rrparenthesis, \gamma} \rangle \wedge \langle \llparenthesis q, \phi  \rrparenthesis, \omega \rangle $}
							
							\item {\changeallfullversiona{$\changenaap{\changenaat{h_3} =} \bigwedge_{\langle p, \gamma \rangle \xrightarrow{ret} \langle q, \epsilon \rangle   \in \Delta}  \langle \changeallfullversionb{p_{\bot}}, \changeallfullversionb{\gamma_{\bot}} \rangle  $}}
							
						\end{itemize}

					}

					\item {\label{BCARET_item:standRulesERG} If $\phi = \changenaaf{E [\phi_1 R^g \phi_2]}$, then, \changenaat{we add to $\Delta'$ the rule:}

						$\langle \llparenthesis p, \phi \rrparenthesis, \gamma \rangle \rightarrow (\langle \llparenthesis p, \phi_2  \rrparenthesis, \gamma \rangle \wedge \langle \llparenthesis p, \phi_1  \rrparenthesis, \gamma \rangle) \vee (\bigvee_{\langle p, \gamma \rangle \xrightarrow{t} \langle q, \omega \rangle \in \Delta}   (\langle \llparenthesis p, \phi_2  \rrparenthesis, \gamma \rangle \wedge  \langle \llparenthesis q, \phi \rrparenthesis, \omega \rangle)$}
					

					\notesctl{The below formula is not fitted in one line. By newline here, we waste one line? Shoud we keep it?}
					
					\item {\label{BCARET_item:standRulesARG} If $\phi = \changenaaf{A [\phi_1 R^g \phi_2]}$, then, \changenaat{we add to $\Delta'$ the rule:}
						
						$\langle \llparenthesis p, \phi \rrparenthesis, \gamma \rangle \rightarrow (\langle \llparenthesis p, \phi_2  \rrparenthesis, \gamma \rangle \wedge \langle \llparenthesis p, \phi_1  \rrparenthesis, \gamma \rangle) \vee (\bigwedge_{\langle p, \gamma \rangle \xrightarrow{t} \langle q, \omega \rangle \in \Delta}   (\langle \llparenthesis p, \phi_2  \rrparenthesis, \gamma \rangle \wedge  \langle \llparenthesis q, \phi \rrparenthesis, \omega \rangle)$}
					

					\item {\label{BCARET_item:standRulesERA} If $\phi = \changenaat{E [\phi_1 R^a \phi_2]}$: $\langle \llparenthesis p, \phi \rrparenthesis, \gamma \rangle \rightarrow (\langle \llparenthesis p, \phi_2  \rrparenthesis, \gamma \rangle \wedge \langle \llparenthesis p, \phi_1  \rrparenthesis, \gamma \rangle) \vee \changenaat{h_1} \vee  \changenaat{h_2} \changeallfullversiona{\vee h_3} \in \Delta'$, \changenaas{where}

						\begin{itemize}

							\item {$\changenaap{\changenaat{h_1} =} \bigvee_{\langle p, \gamma \rangle \xrightarrow{call} \langle q, \gamma_1 \gamma_2 \rangle \in \Delta}  \langle \llparenthesis p, \phi_2  \rrparenthesis, \gamma \rangle \wedge   \langle \changenaad{q},  \gamma_1 \llparenthesis \gamma_2, \phi \rrparenthesis \rangle$}
							\item {$\changenaap{\changenaat{h_2} =} \bigvee_{\langle p, \gamma \rangle \xrightarrow{int} \langle q, \omega\rangle   \in \Delta}  \langle \llparenthesis p, \phi_2  \rrparenthesis, \gamma \rangle \wedge \langle \llparenthesis q, \phi  \rrparenthesis, \omega \rangle $}

							\item {\changeallfullversiona{$\changenaap{\changenaat{h_3} =} \bigvee_{\langle p, \gamma \rangle \xrightarrow{ret} \langle q, \epsilon \rangle   \in \Delta}  \langle \changeallfullversionb{p_{\bot}}, \changeallfullversionb{\gamma_{\bot}} \rangle  $}}
						\end{itemize}

					}

					\item {\label{BCARET_item:standRulesARA} If $\phi = \changenaat{A [\phi_1 R^a \phi_2]}$, $\langle \llparenthesis p, \phi \rrparenthesis, \gamma \rangle \rightarrow (\langle \llparenthesis p, \phi_2  \rrparenthesis, \gamma \rangle \wedge \langle \llparenthesis p, \phi_1  \rrparenthesis, \gamma \rangle) \vee \changeallfullversiona{(\changenaat{h_1} \wedge  \changenaat{h_2} \wedge h_3)} \in \Delta'$, \changenaas{where}

						\begin{itemize}

							\item {$\changenaap{\changenaat{h_1} =} \bigwedge_{\langle p, \gamma \rangle \xrightarrow{call} \langle q, \gamma_1 \gamma_2 \rangle \in \Delta}  \langle \llparenthesis p, \phi_2  \rrparenthesis, \gamma \rangle \wedge   \langle \changenaad{q},  \gamma_1 \llparenthesis \gamma_2, \phi \rrparenthesis \rangle$}
							\item {$\changenaap{\changenaat{h_2} =} \bigwedge_{\langle p, \gamma \rangle \xrightarrow{int} \langle q, \omega\rangle   \in \Delta}  \langle \llparenthesis p, \phi_2  \rrparenthesis, \gamma \rangle \wedge \langle \llparenthesis q, \phi  \rrparenthesis, \omega \rangle $}
							
							\item {\changeallfullversiona{$\changenaap{\changenaat{h_3} =} \bigwedge_{\langle p, \gamma \rangle \xrightarrow{ret} \langle q, \epsilon \rangle   \in \Delta}  \langle \changeallfullversionb{p_{\bot}}, \changeallfullversionb{\gamma_{\bot}} \rangle  $}}
							
						\end{itemize}

					}

					%
					%
					%
					%

					\item {\label{BCARET_item:standRulesValidateAtReturns} \changenaaf{for every $\langle p, \gamma \rangle \xrightarrow{ret} \langle q, \epsilon \rangle \in \Delta$: }

						\begin{itemize}
							
							\item {	\changenaaf{$\langle q, \llparenthesis \gamma'', \phi_1 \rrparenthesis \rangle \rightarrow  \langle \llparenthesis q,  \phi_1 \rrparenthesis,  \gamma'' \rangle  \in \Delta'$ for every $\gamma'' \in \Gamma$,  $\phi_1 \in Cl(\varphi)$ }}

						\end{itemize}

					}

					\item {\label{BCARET_item:standRulesLoopToTrapAbstractOfReturn} \changeallfullversiona{$\langle  \changeallfullversionb{p_{\bot}}, \changeallfullversionb{\gamma_{\bot}} \rangle \rightarrow  \langle  \changeallfullversionb{p_{\bot}}, \changeallfullversionb{\gamma_{\bot}} \rangle \in \Delta'$ }

					}

				\end{enumerate}

				\medskip
				\noindent
				Roughly speaking, the ABPDS $\mathcal{BP}_\varphi$ is a kind of product between $\mathcal{P}$ and the BCARET formula $\varphi$ which ensures that $\mathcal{BP}_\varphi$ has an accepting run from $\langle \llparenthesis p, \varphi \rrparenthesis, \omega \rangle$ iff the configuration $\langle p, \omega \rangle$ satisfies $\varphi$. The form of the control locations of $\mathcal{BP}_\varphi$ is $\llparenthesis p, \phi \rrparenthesis$ where $\phi \in Cl(\varphi)$. Let us explain the intuition behind our construction:

				\begin{itemize} 
					\item {If $\phi = e \in AP$, then, for every $\omega \in \Gamma^*$, $\langle p, \omega \rangle \vDash_{\lambda_f} \phi$ iff $p \in f(e)$. In other words, $\mathcal{BP}_\varphi$ should have an accepting run from $\langle \llparenthesis p, e \rrparenthesis, \omega \rangle$ iff $p \in f(e)$. This is ensured by the transition rules in \ref{BCARET_item:standRulesAP} which add a loop at  $\langle \llparenthesis p, e \rrparenthesis, \omega \rangle$ where $p \in f(e)$ and the fact that $\llparenthesis p, e \rrparenthesis \in F$.}
					
					\item {If $\phi = \neg e \; (e \in AP)$, then, for every $\omega \in \Gamma^*$, $\langle p, \omega \rangle \vDash_{\lambda_f} \phi$ iff $p \notin f(e)$. In other words, $\mathcal{BP}_\varphi$ should have an accepting run from $\langle \llparenthesis p, \neg e \rrparenthesis, \omega \rangle$ iff $p \notin f(e)$. This is ensured by the transition rules in \ref{BCARET_item:standRulesNegAP} which add a loop at  $\langle \llparenthesis p, \neg e \rrparenthesis, \omega \rangle$ where $p \notin f(e)$ and the fact that $\llparenthesis p, \neg e \rrparenthesis \in F$.}

					\item {If $\phi = \phi_1 \wedge \phi_2$, then, for every $\omega \in \Gamma^*$, $\langle p, \omega \rangle \vDash_{\lambda_f} \phi$ iff  ($\langle p, \omega \rangle   \vDash_{\lambda_f} \phi_1$ and  $\langle p, \omega \rangle \vDash_{\lambda_f} \phi_2$). This is ensured by the transition rules in \ref{BCARET_item:standRulesAnd} stating that $\mathcal{BP}_\varphi$ has an accepting run from $\langle \llparenthesis p, \phi_1 \wedge \phi_2 \rrparenthesis, \omega \rangle$ iff $\mathcal{BP}_\varphi$ has an accepting run from both $\langle \llparenthesis p, \phi_1  \rrparenthesis, \omega \rangle$ and  $\langle \llparenthesis p, \phi_2 \rrparenthesis, \omega \rangle$. \ref{BCARET_item:standRulesOr} is similar to \ref{BCARET_item:standRulesAnd}.}

					\item {If $\phi = E [\phi_1 U^g \phi_2]$, then, for every $\omega \in \Gamma^*$, $\langle p, \omega \rangle \vDash_{\lambda_f} \phi$ iff  $\langle p, \omega \rangle \vDash_{\lambda_f} \phi_2$ or  ($\langle p, \omega \rangle \vDash_{\lambda_f} \phi_1$ and there exists an immediate successor $\langle p', \omega' \rangle$ of $\langle p, \omega \rangle$ s.t. $\langle p', \omega' \rangle \vDash_{\lambda_f} \phi$). This is ensured by the transition rules in \ref{BCARET_item:standRulesEUG} stating that $\mathcal{BP}_\varphi$ has an accepting run from $\langle \llparenthesis p, E [\phi_1 U^g \phi_2] \rrparenthesis, \omega \rangle$ iff $\mathcal{BP}_\varphi$ has an accepting run from $\langle \llparenthesis p, \phi_2  \rrparenthesis, \omega \rangle$ or  ($\mathcal{BP}_\varphi$ has an accepting run from both  $\langle \llparenthesis p, \phi_1  \rrparenthesis, \omega \rangle$ and  $\langle \llparenthesis p', \phi   \rrparenthesis, \omega' \rangle$ where   $\langle p', \omega' \rangle$ is an immediate successor of $\langle p, \omega \rangle$). \ref{BCARET_item:standRulesAUG} is similar to \ref{BCARET_item:standRulesEUG}. }

					\item {If $\phi = E [\phi_1 R^g \phi_2]$, then, for every $\omega \in \Gamma^*$, $\langle p, \omega \rangle \vDash_{\lambda_f} \phi$ iff  ($\langle p, \omega \rangle \vDash_{\lambda_f} \phi_2$ and $\langle p, \omega \rangle \vDash_{\lambda_f} \phi_1$) or ($\langle p, \omega \rangle \vDash_{\lambda_f} \phi_2$ and there exists an immediate successor $\langle p', \omega' \rangle$ of $\langle p, \omega \rangle$ s.t. $\langle p', \omega' \rangle \vDash_{\lambda_f} \phi$). This is ensured by the transition rules in \ref{BCARET_item:standRulesERG} stating that $\mathcal{BP}_\varphi$ has an accepting run from $\langle \llparenthesis p, E [\phi_1 R^g \phi_2] \rrparenthesis, \omega \rangle$ iff $\mathcal{BP}_\varphi$ has an accepting run from both $\langle \llparenthesis p, \phi_2  \rrparenthesis, \omega \rangle$ and $\langle \llparenthesis p, \phi_1  \rrparenthesis, \omega \rangle$; or $\mathcal{BP}_\varphi$ has an accepting run from both $\langle \llparenthesis p, \phi_2  \rrparenthesis, \omega \rangle$ and  $\llparenthesis p', \phi   \rrparenthesis, \omega' \rangle$ where  $\langle p', \omega' \rangle$ is an immediate successor of $\langle p, \omega \rangle$. In addition, for $R^g$ formulas, the \textit{stop} condition is not required, i.e, for a formula $\phi_1 R^g \phi_2$ that is applied to a specific run, we don't require that $\phi_1$ must eventually hold. To ensure that the runs on which $\phi_2$ always holds are accepted, we add $\llparenthesis p, \phi \rrparenthesis$ to the B\"{u}chi accepting condition $F$ (via the subset $F_3$ of $F$). \ref{BCARET_item:standRulesARG} is similar to \ref{BCARET_item:standRulesERG}. }

					\item {If $\phi = EX^g \phi_1$, then, for every $\omega \in \Gamma^*$, $\langle p, \omega \rangle \vDash_{\lambda_f} \phi$ iff there exists an immediate successor $\langle p', \omega' \rangle$ of $\langle p, \omega \rangle$ s.t. $\langle p', \omega' \rangle \vDash_{\lambda_f} \phi_1$. This is ensured by the transition rules in \ref{BCARET_item:standRulesEXG} stating that $\mathcal{BP}_\varphi$ has an accepting run from $\langle \llparenthesis p, EX^g \phi_1 \rrparenthesis, \omega \rangle$ iff there exists an immediate successor $\langle p', \omega' \rangle$ of $\langle p, \omega \rangle$ s.t. $\mathcal{BP}_\varphi$ has an accepting run from $\langle \llparenthesis p', \phi_1  \rrparenthesis, \omega' \rangle$. \ref{BCARET_item:standRulesAXG} is similar to \ref{BCARET_item:standRulesEXG}. }

					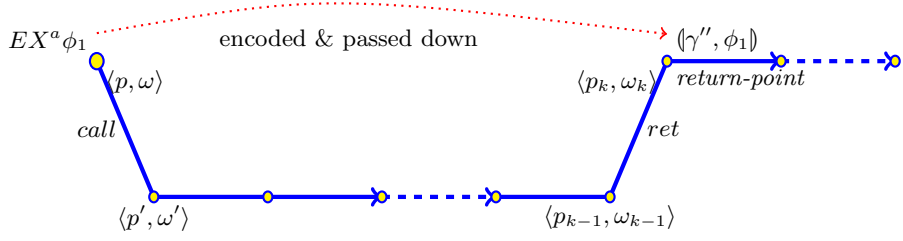
\begin{figure*}[ht]
						\centering
						
						\begin{tikzpicture}[xscale=1.5, yscale=1.8]
						\tikzset{lineStyle/.style={blue, ultra thick}};
						\tikzset{myDot/.style={blue, fill = yellow, thick}};
						\tikzset{myRectangleNode/.style={rectangle, thick, draw= black, below right, black}};
						
						
						\draw [->, lineStyle]  (2,0) -- (2.5, -1) -- (3.5, -1) -- (4.5, -1);

						\draw [->, lineStyle, dashed] (4.5, -1) -- (5.5, -1);
						
						\draw [->, lineStyle] (5.5, -1) -- (6.5, -1) -- (7, 0) -- (8, 0);
						\draw [->, lineStyle, dashed] (8, 0) -- (9, 0);

						\draw [myDot] (2,0) circle [radius=0.06];
						\draw [myDot] (2.5, -1) circle [radius=0.04];
						\draw [myDot] (3.5, -1) circle [radius=0.04];
						\draw [myDot] (4.5, -1) circle [radius=0.04];
						\draw [myDot] (5.5, -1) circle [radius=0.04];
						\draw [myDot] (6.5, -1) circle [radius=0.04];
						\draw [myDot] (7,0) circle [radius=0.04];
						\draw [myDot] (8,0) circle [radius=0.04];
						\draw [myDot] (9,0) circle [radius=0.04];
						
						\node[ left](call) at (2.25,-0.5) {$call$};

						\node[above left](AtomA_call) at (2,0) { $EX^a \phi_1$ };

						\node[right](ret) at (6.75,-0.5) {$ret$};

						\node[below right](return-point) at (7,0) {\textit{return-point}};
						\node[above right](AtomA0_pass) at (7,0) {$\llparenthesis \gamma'', \phi_1 \rrparenthesis$};
						
						
						\draw[dotted, red, thick, ->] (AtomA_call).. controls(4,0.5).. (AtomA0_pass);
						
						\node[below right](encoded) at (3,0.3) {encoded $\&$ passed down};

						\node[below right]() at (2,0) {$\langle p, \omega \rangle$};
						\node[below]() at (2.5,-1) {$\langle p', \omega' \rangle$};
						\node[below]() at (6.5,-1) {$\langle p_{k-1}, \omega_{k-1} \rangle$};
						\node[below left]() at (7,0) {$\langle p_k, \omega_k \rangle$};
						
						\end{tikzpicture}
						
						\caption{$\langle p, \omega \rangle$ $\xRightarrow{}_{\mathcal{P}}$ $\langle p', \omega' \rangle$ corresponds to a call statement} 
						\label{BCARET_fig:nextAbstract}
					\end{figure*}

					\item {If $\phi = EX^a \phi_1$, then,  for every $\omega \in \Gamma^*$, $\langle p, \omega \rangle \vDash_{\lambda_f} \phi$ iff there exists an abstract-successor $\langle p_k, \omega_k \rangle$ of $\langle p, \omega \rangle$ s.t. $\langle p_k, \omega_k \rangle \vDash_{\lambda_f} \phi_1$ \myIntuitionLabel{intuitionEXAFact}. Let $\uppi \in Traces(\langle p, \omega \rangle)$ be a run starting from $\langle p, \omega \rangle$ on which  $\langle p_k, \omega_k \rangle$ is the abstract-successor of $\langle p, \omega \rangle$.  Over $\uppi$, let $\langle p', \omega' \rangle$ be the immediate successor of $\langle p, \omega \rangle$. In what follows, we explain how we can ensure \refMyIntuitionLabel{intuitionEXAFact}.}

					\medskip
					\noindent
					\textbf{1.} Firstly, we show that for every abstract-successor $\langle p_k, \omega_k \rangle \neq \bot$  of $\langle p, \omega \rangle$,  $ \langle \llparenthesis p, EX^a \phi_1 \rrparenthesis, \omega \rangle \xRightarrow{}_{\mathcal{BP}_\varphi} \langle \llparenthesis p_k, \phi_1  \rrparenthesis, \omega_k \rangle $. There are two possibilities:

					\begin{itemize}

						\item
						{If $\langle p, \omega \rangle$ $\xRightarrow{}_{\mathcal{P}}$ $\langle p', \omega' \rangle$ corresponds to a call statement.
							Let us consider Figure \ref{BCARET_fig:nextAbstract} to explain this case. $\langle \llparenthesis p, \phi \rrparenthesis, \omega \rangle \xRightarrow{}_{\mathcal{BP}_\varphi} \langle \llparenthesis p_k, \phi_1 \rrparenthesis, \omega_k \rangle$ is ensured by rules corresponding to $h_1$ in \ref{BCARET_item:standRulesEXA}, the rules in $\Delta \subseteq \Delta'$ and the rules in \ref{BCARET_item:standRulesValidateAtReturns} as follows: rules corresponding to $h_1$ in \ref{BCARET_item:standRulesEXA} allow to record $\phi_1$ in the return point of the call, rules in $\Delta \subseteq \Delta'$ allow to mimic the run of the PDS $\mathcal{P}$ and rules in \ref{BCARET_item:standRulesValidateAtReturns} allow to extract and put back $\phi_1$ when the return-point is reached. In what follows, we show in more details how this works:
							Let $\langle p, \gamma \rangle \xrightarrow{call} \langle p', \gamma' \gamma'' \rangle$ be the rule associated with the transition $\langle p, \omega \rangle$ $\xRightarrow{}_{\mathcal{P}}$ $\langle p', \omega' \rangle$, then we have $\omega = \gamma \omega''$ and $\omega' = \gamma' \gamma'' \omega''$. Let $\langle p_{k-1}, \omega_{k-1} \rangle$ $\xRightarrow{}_{\mathcal{P}}$ $\langle p_{k}, \omega_{k} \rangle$ be the transition that corresponds to the $ret$ statement of this call on $\uppi$.	Let then $\langle p_{k-1}, \beta \rangle \xrightarrow{ret} \langle p_k, \epsilon \rangle \in \Delta$ be the corresponding return rule. Then, we have necessarily $\omega_{k-1} = \beta \gamma'' \omega''$, \ifdefined \ThesisVersion since as explained in Section \ref{thesis_sec:gammaTwoTopOfStackAtReturn} \else since as explained in Section \ref{BCARET_sec:PDS}\fi,
							$\gamma''$ is the return address of the call. After applying this rule, $\omega_{k} = \gamma'' \omega''$.
							In other words, $\gamma''$ will be the topmost stack symbol at the corresponding return point of the call. So, in order to ensure that $\langle \llparenthesis p, \phi \rrparenthesis, \omega \rangle \xRightarrow{}_{\mathcal{BP}_\varphi} \langle \llparenthesis p_k, \phi_1 \rrparenthesis, \omega_k \rangle$, we proceed as follows:
							At the call $\langle p, \gamma \rangle \xrightarrow{call} \langle p', \gamma' \gamma'' \rangle$, we encode the formula $\phi_1$ into $\gamma''$ by the rule corresponding to $h_1$ in \ref{BCARET_item:standRulesEXA} stating that $\langle \llparenthesis p, EX^a \phi_1 \rrparenthesis,  \gamma \rangle \xrightarrow{} \langle p', \gamma' 
							\llparenthesis \gamma'', \phi_1 \rrparenthesis \rangle \allowbreak \in \Delta'$. 
							This allows to record $\phi_1$ in the corresponding return point of the stack. After that, the rules in $\Delta \subseteq \Delta'$ allow $\mathcal{BP}_\varphi$ to mimic the run $\uppi$ of $\mathcal{P}$ from $\langle p', \omega' \rangle$ till the corresponding return-point of this call, where $\llparenthesis \gamma'', \phi_1 \rrparenthesis$ is the topmost stack symbol. More specifically, the following sequence of $\mathcal{P}$: $\langle p', \gamma' \gamma'' \omega'' \rangle \xRightarrow{*}_{\mathcal{P}} \langle p_{k-1}, \beta \gamma'' \omega'' \rangle \xRightarrow{*}_{\mathcal{P}} \langle p_{k}, \gamma'' \omega'' \rangle  $  will be mimicked by the following sequence of $\mathcal{BP}_\varphi$: $ \langle \llparenthesis p', \gamma' \llparenthesis \gamma'', \phi_1 \rrparenthesis  \omega'' \rangle \allowbreak \xRightarrow{}_{\mathcal{BP}_\varphi}  \langle p_{k-1}, \beta \llparenthesis \gamma'', \phi_1 \rrparenthesis \omega'' \rangle \allowbreak \xRightarrow{}_{\mathcal{BP}_\varphi} \langle p_{k},  \llparenthesis \gamma'', \phi_1 \rrparenthesis \omega'' \rangle$ using the rules of $\Delta$. At the return-point, we extract $\phi_1$ from the stack and encode it into $p_k$ by adding the transition rules in \ref{BCARET_item:standRulesValidateAtReturns} $\langle p_k, \llparenthesis \gamma'', \phi_1 \rrparenthesis \rangle \rightarrow  \langle \llparenthesis p_k,  \phi_1 \rrparenthesis,  \gamma'' \rangle$. Therefore, we obtain that $\langle \llparenthesis p, \phi \rrparenthesis, \omega \rangle \xRightarrow{}_{\mathcal{BP}_\varphi} \langle \llparenthesis p_k, \phi_1 \rrparenthesis, \omega_k \rangle$. The property holds for this case.

							\item{If $\langle p, \omega \rangle$ $\xRightarrow{}_{\mathcal{P}}$ $\langle p', \omega' \rangle$ corresponds to a simple statement. Then, the abstract successor of $\langle p, \omega \rangle$ is its immediate successor $\langle p', \omega' \rangle$. Thus, we get that $\langle p_k, \omega_k \rangle = \langle p', \omega' \rangle$.  From the transition rules corresponding to $h_2$ in \ref{BCARET_item:standRulesEXA}, we get that $ \langle \llparenthesis p, EX^a \phi_1 \rrparenthesis, \omega \rangle \xRightarrow{}_{\mathcal{BP}_\varphi} \langle \llparenthesis p', \phi_1  \rrparenthesis, \omega' \rangle $. Therefore, $ \langle \llparenthesis p, EX^a \phi_1 \rrparenthesis, \omega \rangle \xRightarrow{}_{\mathcal{BP}_\varphi} \langle \llparenthesis p_k, \phi_1  \rrparenthesis, \omega_k \rangle $. The property holds for this case.}

						}
						
					\end{itemize}

					\noindent
					\textbf{2.} Now, let us consider the case where $\langle p_k, \omega_k \rangle$, the abstract successor of $\langle p, \omega \rangle$, is $\bot$. This case occurs when $\langle p, \omega \rangle$ $\xRightarrow{}_{\mathcal{P}}$ $\langle p', \omega' \rangle$ corresponds to a return statement. Then, one abstract successor of $\langle p, \omega \rangle$ is $\bot$. Note that $\bot$ does not satisfy any formula, i.e., $\bot$  does not satisfy $\phi_1$. Therefore, from $ \langle \llparenthesis p,  EX^a \phi_1  \rrparenthesis, \omega \rangle$, we need to ensure that the path of $\mathcal{BP}_\varphi$ reflecting the possibility in \refMyIntuitionLabel{intuitionEXAFact} that $\langle p_k, \omega_k \rangle \vDash_{\lambda_f} \phi_1$ is not accepted. To do this, we exploit additional trap configurations. We use $p_{\bot}$ and $\gamma_{\bot}$ as trap control location and trap stack symbol to obtain these trap configurations. To be more specific, let $\langle p, \gamma \rangle \xrightarrow{ret} \langle p',  \epsilon \rangle$ be the rule associated with the transition $\langle p, \omega \rangle$ $\xRightarrow{}_{\mathcal{P}}$ $\langle p', \omega' \rangle$, then we have $\omega = \gamma \omega''$ and $\omega' = \omega''$. We add the transition rule corresponding to $h_3$ in \ref{BCARET_item:standRulesEXA} to allow $ \langle \llparenthesis p,  EX^a \phi_1  \rrparenthesis, \omega \rangle \xRightarrow{}_{\mathcal{BP}_\varphi} \langle p_{\bot},   \gamma_{\bot} \omega'' \rangle $. Since a run of $\mathcal{BP}_\varphi$  includes only infinite paths, we equip these trap configurations with self-loops by the transition rules in \ref{BCARET_item:standRulesLoopToTrapAbstractOfReturn}, i.e., $\langle p_{\bot},   \gamma_{\bot} \omega'' \rangle  \xRightarrow{}_{\mathcal{BP}_\varphi}  \langle p_{\bot},   \gamma_{\bot} \omega'' \rangle $.  As a result, we obtain a corresponding path in $\mathcal{BP}_\varphi$:  $ \langle \llparenthesis p,  EX^a \phi_1  \rrparenthesis, \omega \rangle \xRightarrow{}_{\mathcal{BP}_\varphi}  \langle p_{\bot},   \gamma_{\bot} \omega'' \rangle  \xRightarrow{}_{\mathcal{BP}_\varphi}  \langle p_{\bot},   \gamma_{\bot} \omega'' \rangle $. Note that this path is not accepted by $\mathcal{BP}_\varphi$ because $p_{\bot} \notin F$.

					\medskip
					\noindent
					In summary, for every abstract-successor $\langle p_k, \omega_k \rangle$  of $\langle p, \omega \rangle$, if $\langle p_k, \omega_k \rangle \neq \bot$, then, $ \langle \llparenthesis p, EX^a \phi_1 \rrparenthesis, \omega \rangle \xRightarrow{}_{\mathcal{BP}_\varphi} \langle \llparenthesis p_k, \phi_1  \rrparenthesis, \omega_k \rangle $; otherwise $ \langle \llparenthesis p,  EX^a \phi_1  \rrparenthesis, \omega \rangle \xRightarrow{}_{\mathcal{BP}_\varphi}  \langle p_{\bot},   \gamma_{\bot} \omega'' \rangle  \xRightarrow{}_{\mathcal{BP}_\varphi}  \langle p_{\bot},   \gamma_{\bot} \omega'' \rangle $ which is not accepted by $\mathcal{BP}_\varphi$. Therefore, \refMyIntuitionLabel{intuitionEXAFact} is ensured by the transition rules in \ref{BCARET_item:standRulesEXA} stating that $\mathcal{BP}_\varphi$ has an accepting run from $\langle \llparenthesis p, EX^a \phi_1 \rrparenthesis, \omega \rangle$ iff there exists an abstract successor $\langle p_k, \omega_k \rangle$ of $\langle p, \omega \rangle$ s.t. $\mathcal{BP}_\varphi$ has an accepting run from $\langle \llparenthesis p_k, \phi_1  \rrparenthesis, \omega_k \rangle$.

					%
					%
					%

					\item {If $\phi = AX^a \phi_1$: this case is ensured by the transition rules in \ref{BCARET_item:standRulesAXA} together with \ref{BCARET_item:standRulesValidateAtReturns} and $\Delta \subseteq \Delta'$. The intuition of \ref{BCARET_item:standRulesAXA} is similar to that of \ref{BCARET_item:standRulesEXA}. }

					\item {If $\phi = E [\phi_1 U^a \phi_2]$, then, for every $\omega \in \Gamma^*$, $\langle p, \omega \rangle \vDash_{\lambda_f} \phi$ iff  $\langle p, \omega \rangle \vDash_{\lambda_f} \phi_2$ or  ($\langle p, \omega \rangle \vDash_{\lambda_f} \phi_1$ and there exists an abstract successor $\langle p_k, \omega_k \rangle$ of $\langle p, \omega \rangle$ s.t. $\langle p_k, \omega_k \rangle \vDash_{\lambda_f} \phi$) \myIntuitionLabel{intuitionEUAFact}. 	Let $\uppi \in Traces(\langle p, \omega \rangle)$ be a run starting from $\langle p, \omega \rangle$ on which  $\langle p_k, \omega_k \rangle$ is the abstract-successor of $\langle p, \omega \rangle$.  Over $\uppi$, let $\langle p', \omega' \rangle$ be the immediate successor of $\langle p, \omega \rangle$. }

					\medskip
					\noindent
					\textbf{1.} Firstly, we show that for every abstract-successor $\langle p_k, \omega_k \rangle \neq \bot$  of $\langle p, \omega \rangle$,  $ \langle \llparenthesis p, \phi  \rrparenthesis, \omega \rangle \xRightarrow{}_{\mathcal{BP}_\varphi} \{ \langle \llparenthesis p, \phi_1  \rrparenthesis, \omega \rangle, \langle \llparenthesis p_k, \phi  \rrparenthesis, \omega_k \rangle \} $. There are two possibilities:

					\begin{itemize}
						\item {If $\langle p, \omega \rangle$ $\xRightarrow{}_{\mathcal{P}}$ $\langle p', \omega' \rangle$ corresponds to a call statement. From the rules corresponding to $h_1$ in \ref{BCARET_item:standRulesEUA}, we get that  $ \langle \llparenthesis p, \phi  \rrparenthesis, \omega \rangle \xRightarrow{}_{\mathcal{BP}_\varphi} \{ \langle \llparenthesis p, \phi_1  \rrparenthesis, \omega \rangle, \langle p', \omega' \rangle \} $ where $\langle p', \omega' \rangle$ is the immediate successor of $\langle p, \omega \rangle$.  Thus, to ensure that $ \langle \llparenthesis p, \phi  \rrparenthesis, \omega \rangle \allowbreak \xRightarrow{}_{\mathcal{BP}_\varphi} \{ \langle \llparenthesis p, \phi_1  \rrparenthesis, \omega \rangle, \langle \llparenthesis p_k, \phi  \rrparenthesis, \omega_k \rangle \} $, we only need to ensure that $\langle p', \omega' \rangle \xRightarrow{}_{\mathcal{BP}_\varphi} \langle \llparenthesis p_k, \phi \rrparenthesis, \omega_k \rangle$. As for the case $\phi = EX^a \phi_1$, $\langle p', \omega' \rangle \xRightarrow{}_{\mathcal{BP}_\varphi} \langle \llparenthesis p_k, \phi \rrparenthesis, \omega_k \rangle$ is ensured by the rules in $\Delta \subseteq \Delta'$ and the rules in \ref{BCARET_item:standRulesValidateAtReturns}: rules in $\Delta \subseteq \Delta'$ allow to mimic the run of the PDS $\mathcal{P}$ before the return and rules in \ref{BCARET_item:standRulesValidateAtReturns} allow to extract and put back $\phi_1$ when the return-point is reached.}

						\item{If $\langle p, \omega \rangle$ $\xRightarrow{}_{\mathcal{P}}$ $\langle p', \omega' \rangle$ corresponds to a simple statement. Then, the abstract successor of $\langle p, \omega \rangle$ is its immediate successor $\langle p', \omega' \rangle$. Thus, we get that $\langle p_k, \omega_k \rangle = \langle p', \omega' \rangle$.  From the transition rules corresponding to $h_2$ in \ref{BCARET_item:standRulesEUA}, we get that $ \langle \llparenthesis p, E [\phi_1 U^a \phi_2] \rrparenthesis, \omega \rangle \xRightarrow{}_{\mathcal{BP}_\varphi} \{ \langle \llparenthesis p, \phi_1  \rrparenthesis, \omega \rangle, \langle \llparenthesis p', \phi  \rrparenthesis, \omega' \rangle \}$. Therefore, $ \langle \llparenthesis p, E [\phi_1 U^a \phi_2] \rrparenthesis, \omega \rangle \xRightarrow{}_{\mathcal{BP}_\varphi} \{ \langle \llparenthesis p, \phi_1  \rrparenthesis, \omega \rangle, \langle \llparenthesis p_k, \phi  \rrparenthesis, \omega_k \rangle \}$. In other words, $\mathcal{BP}_\varphi$ has an accepting run from both  $\langle \llparenthesis p, \phi_1  \rrparenthesis, \omega \rangle$ and  $\langle \llparenthesis p_k, \phi   \rrparenthesis, \omega_k \rangle$ where   $\langle p_k, \omega_k \rangle$ is an abstract successor of $\langle p, \omega \rangle$. The property holds for this case.}
					\end{itemize}


					\medskip
					\noindent
					\textbf{2.} Now, let us consider the case where $\langle p_k, \omega_k \rangle = \bot$. As explained previously, this case occurs when $\langle p, \omega \rangle$ $\xRightarrow{}_{\mathcal{P}}$ $\langle p', \omega' \rangle$ corresponds to a return statement. Then, the abstract successor of $\langle p, \omega \rangle$ is $\bot$. Note that $\bot$ does not satisfy any formula, i.e., $\bot$  does not satisfy $\phi$. Therefore, from $ \langle \llparenthesis p,   E [\phi_1 U^a \phi_2] \rrparenthesis, \omega \rangle$, we need to ensure that the path reflecting the possibility in \refMyIntuitionLabel{intuitionEUAFact}  that ($\langle p, \omega \rangle \vDash_{\lambda_f} \phi_1$ and  $\langle p_k, \omega_k \rangle \vDash_{\lambda_f} \phi$) is not accepted by $\mathcal{BP}_\varphi$. This is ensured as for the case $\phi = EX^a \phi_1$ by the transition rules corresponding to $h_3$ in \ref{BCARET_item:standRulesEUA}.

					\medskip
					\noindent 
					In summary, for every abstract-successor $\langle p_k, \omega_k \rangle$  of $\langle p, \omega \rangle$, if $\langle p_k, \omega_k \rangle \neq \bot$, then, $ \langle \llparenthesis p, E [\phi_1 U^a \phi_2] \rrparenthesis, \omega \rangle \xRightarrow{}_{\mathcal{BP}_\varphi}  \{ \langle \llparenthesis p, \phi_1  \rrparenthesis, \omega \rangle, \langle \llparenthesis p_k, E [\phi_1 U^a \phi_2]  \rrparenthesis, \omega_k \rangle \} $; otherwise $ \langle \llparenthesis p,  E [\phi_1 U^a \phi_2] \rrparenthesis, \omega \rangle \xRightarrow{}_{\mathcal{BP}_\varphi}  \langle p_{\bot},   \gamma_{\bot} \omega'' \rangle  \xRightarrow{}_{\mathcal{BP}_\varphi}  \langle p_{\bot},   \gamma_{\bot} \omega'' \rangle $ which is not accepted by $\mathcal{BP}_\varphi$. Therefore, \refMyIntuitionLabel{intuitionEUAFact} is ensured by the transition rules in \ref{BCARET_item:standRulesEUA} stating that  $\mathcal{BP}_\varphi$ has an accepting run from $\langle \llparenthesis p, E [\phi_1 U^a \phi_2] \rrparenthesis, \omega \rangle$ iff  $\mathcal{BP}_\varphi$ has an accepting run from $\langle \llparenthesis p, \phi_2  \rrparenthesis, \omega \rangle$; or $\mathcal{BP}_\varphi$ has an accepting run from both  $\langle \llparenthesis p, \phi_1  \rrparenthesis, \omega \rangle$ and  $\langle \llparenthesis p_k, E [\phi_1 U^a \phi_2]   \rrparenthesis, \omega_k \rangle$ where   $\langle p_k, \omega_k \rangle$ is an abstract successor of $\langle p, \omega \rangle$.


					\item {The intuition behind the rules corresponding to the cases $\phi = A [\phi_1 U^a \phi_2]$, \changeifma{$\phi = E [\phi_1 R^a \phi_2]$}, $\phi = A [\phi_1 R^a \phi_2]$ are similar to the \changeifma{previous cases}.}

				\end{itemize}

				\medskip
				\noindent
				\textbf{The B\"{u}chi accepting condition.} The elements of the B\"{u}chi accepting condition set $F$ of $\mathcal{BP}_{\varphi}$  ensure the liveness requirements of until-formulas on infinite global paths, infinite abstract paths as well as on finite abstract paths. 
				
				\begin{mycolorforparablue}
					\begin{itemize}
						
						\item { With regards to infinite global paths, the fact that the liveness requirement $\phi_{2}$ in $E[\phi_{1} U^g \phi_{2}]$ is eventually satisfied in $\mathcal{P}$ is ensured by the fact that $\llparenthesis p, E[\phi_{1} U^g \phi_{2}] \rrparenthesis$ doesn't belong to $F$.  Note that $ \langle p, \omega \rangle \vDash_{\lambda_f}  E[\phi_{1} U^g \phi_{2}]$ iff $ \langle p, \omega \rangle \vDash_{\lambda_f} \phi_2$  or there exists a global-successor $\langle p', \omega' \rangle$ s.t. ($ \langle p, \omega \rangle \vDash_{\lambda_f}  \phi_{1} $  and $ \langle p', \omega' \rangle \vDash_{\lambda_f}  E[\phi_{1} U^g \phi_{2}]$). Because $\phi_2$ should hold eventually, to avoid the case where a run of $\mathcal{BP_{\varphi}}$ always carries $E[\phi_{1} U^g \phi_{2}]$ and never reaches $\phi_2$, we don't set $\llparenthesis p, E[\phi_{1} U^g \phi_{2}] \rrparenthesis$  as an element of the B\"{u}chi accepting condition set. This guarantees that the accepting run of $\mathcal{BP}_{\varphi}$ must visit some control locations in $\llparenthesis p,  \phi_{2} \rrparenthesis$ which ensures that $\phi_2$ will eventually hold. The liveness requirements of $A[\phi_{1} U^g \phi_{2}]$ are ensured as for the case of $E[\phi_{1} U^g \phi_{2}]$.}
						
						\item { With regards to infinite abstract paths, the fact that the liveness requirement $\phi_{2}$ in $E[\phi_{1} U^a \phi_{2}]$ is eventually satisfied in $\mathcal{P}$ is ensured by the fact that $\llparenthesis p, E[\phi_{1} U^a \phi_{2}] \rrparenthesis$ doesn't belong to $F$. The intuition behind this case is similar to the intuition of $E[\phi_{1} U^g \phi_{2}]$. The liveness requirements of $A[\phi_{1} U^a \phi_{2}]$ are ensured as for the case of $E[\phi_{1} U^a \phi_{2}]$.}

						\begin{figure*}
							\centering
							\begin{tikzpicture}[xscale=1.5, yscale=1.5]
							\tikzset{lineStyle/.style={blue, ultra thick}};
							\tikzset{myDot/.style={blue, fill = yellow, thick}};
							\tikzset{myRectangleNode/.style={rectangle, thick, draw= black, below right, black}};
							\draw [->, lineStyle, dashed] (0, 0) -- (1,0);
							\draw [->, lineStyle] (1,0) -- (2,0) -- (2.5, -1) -- (3.5, -1) -- (4.5, -1);
							\draw [->, lineStyle, dashed] (4.5, -1) -- (5.5, -1);
							\draw [->, lineStyle] (5.5, -1) -- (6.5, -1) -- (7, 0) -- (8, 0);
							\draw [->, lineStyle, dashed] (8, 0) -- (9, 0);
							\draw [myDot] (0,0) circle [radius=0.04];
							\draw [myDot] (1,0) circle [radius=0.04];
							\draw [myDot] (2,0) circle [radius=0.04];
							\draw [myDot] (2.5, -1) circle [radius=0.04];
							\draw [myDot] (3.5, -1) circle [radius=0.04];
							\draw [myDot] (4.5, -1) circle [radius=0.04];
							\draw [myDot] (5.5, -1) circle [radius=0.04];
							\draw [myDot] (6.5, -1) circle [radius=0.04];
							\draw [myDot] (7,0) circle [radius=0.04];
							\draw [myDot] (8,0) circle [radius=0.04];
							\draw [myDot] (9,0) circle [radius=0.04];
							\node[below left](call) at (2,0) {$call$};
							
							\node[above] at (0,0) {$ $};
							\node[above right](AtomA_call) at (2,0) {$EX^a \phi_1$};
							\node[above right]() at (2.5,-1) {$ $};
							
							\node[above left](proc-entry) at (2.3,-1) {$proc$};
							\node[right](ret) at (6.5,-1) {$ret$};
							\node[below right](return-point) at (7,0) {\textit{return-point}};
							\node[above left](AtomA0_pass) at (7,0) {$\llparenthesis \gamma'', \phi_1 \rrparenthesis$};
							\draw[dotted, red, thick, ->] (AtomA_call).. controls(4,0.5).. (AtomA0_pass);
							\node[below right](encoded) at (3,0.3) {encoded $\&$ passed down};
							\node[below]() at (0,0) {$\langle p_0, \omega_0 \rangle$};
							\node[below right]() at (2,0) {$\langle p_i, \omega_i \rangle$};
							\node[below]() at (2.5,-1) {$\langle p_{i+1}, \omega_{i+1} \rangle$};

							\node[above]() at (5.5,-1) {$\langle p_{k-2}, \omega_{k-2} \rangle$};
							\node[below]() at (6.5,-1) {$\langle p_{k-1}, \omega_{k-1} \rangle$};
							\node[above right]() at (7,0) {$\langle p_k, \omega_k \rangle$};
							\end{tikzpicture}
							\caption{$\langle p_{i}, \omega_{i} \rangle$ finally reach its corresponding return-point}
							\label{BCARET_fig:nextAbstractToShowFinitePath}
						\end{figure*}
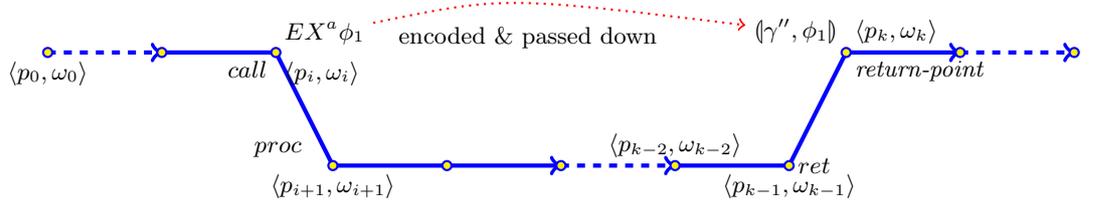

						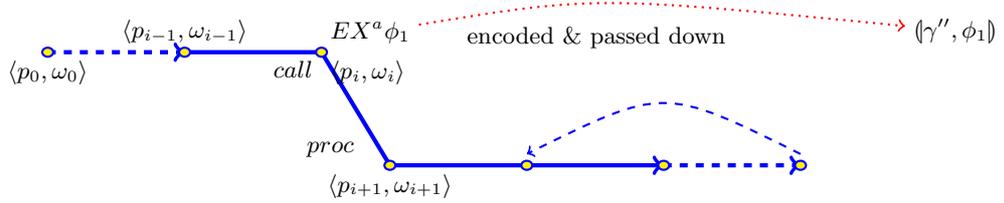
\begin{figure*}
							\centering
							\begin{tikzpicture}[xscale=1.8, yscale=1.5]
							\tikzset{lineStyle/.style={blue, ultra thick}};
							\tikzset{myDot/.style={blue, fill = yellow, thick}};
							\tikzset{myRectangleNode/.style={rectangle, thick, draw= black, below right, black}};
							\draw [->, lineStyle, dashed] (0, 0) -- (1,0);
							\draw [->, lineStyle] (1,0) -- (2,0) -- (2.5, -1) -- (3.5, -1) -- (4.5, -1);
							\draw [->, lineStyle, dashed] (4.5, -1) -- (5.5, -1);


							\draw [myDot] (0,0) circle [radius=0.04];
							\draw [myDot] (1,0) circle [radius=0.04];
							\draw [myDot] (2,0) circle [radius=0.04];
							\draw [myDot] (2.5, -1) circle [radius=0.04];
							\draw [myDot] (3.5, -1) circle [radius=0.04];
							\draw [myDot] (4.5, -1) circle [radius=0.04];
							\draw [myDot] (5.5, -1) circle [radius=0.04];
							\node[below left](call) at (2,0) {$call$};
							
							\node[above] at (0,0) {$ $};
							\node[above right](AtomA_call) at (2,0) {$EX^a \phi_1$};
							\node[above right]() at (2.5,-1) {$ $};

							\node[above left](proc-entry) at (2.3,-1) {$proc$};
							
							\node[above left](AtomA0_pass) at (7,0) {$\llparenthesis \gamma'', \phi_1 \rrparenthesis$};
							
							\draw[dotted, red, thick, ->] (AtomA_call).. controls(4,0.5).. (AtomA0_pass);

							\draw[dashed, blue, thick, ->] (5.5,-.9).. controls(4.5, -.3).. (3.5,-.9);

							\node[below right](encoded) at (3,0.3) {encoded $\&$ passed down};
							\node[below]() at (0,0) {$\langle p_0, \omega_0 \rangle$};
							\node[below right]() at (2,0) {$\langle p_i, \omega_i \rangle$};
							\node[below]() at (2.5,-1) {$\langle p_{i+1}, \omega_{i+1} \rangle$};

							\node[above]() at (1,0) {$\langle p_{i-1}, \omega_{i-1} \rangle$};
							\node[below]() at (1,0) {$ $};	
							
							
							\end{tikzpicture}
							\caption{$\langle p_{i}, \omega_{i} \rangle$ never reach its corresponding return-point}
							\label{BCARET_fig:nextAbstractWithoutMatching}
						\end{figure*}

						\item {With regards to finite abstract paths $\langle p_0, \omega_0 \rangle \langle p_1, \omega_1 \rangle... \langle p_m, \omega_m \rangle$ where $\langle p_{m}, \omega_{m} \rangle \allowbreak \xRightarrow{}_{\mathcal{P}} \langle p_{m+1}, \omega_{m+1} \rangle$ corresponds to a return statement, the fact that the liveness requirement $\phi_{2}$ in $E[\phi_{1} U^g \phi_{2}]$ is eventually satisfied in $\mathcal{P}$ is ensured by the fact that $p_{\bot}$ doesn't belong to $F$.  Look at Figure \ref{BCARET_fig:nextAbstractToShowFinitePath} for an illustration. In this figure, for every $i + 1 \leq u \leq k-1$, the abstract path starting from $\langle p_u, \omega_u \rangle$ is finite because the abstract successor of $\langle p_{k-1}, \omega_{k-1} \rangle$ is $\bot$ since $\langle p_{k-1}, \omega_{k-1} \rangle \xRightarrow{}_{\mathcal{P}} \langle p_{k}, \omega_{k} \rangle$ corresponds to a return statement. Suppose that we want to check whether  $ \langle p_{k-1}, \omega_{k-1} \rangle \vDash_{\lambda_f}  E[\phi_{1} U^a \phi_{2}]$, then, we get that $ \langle p_{k-1}, \omega_{k-1} \rangle \vDash_{\lambda_f}  E[\phi_{1} U^a \phi_{2}]$ iff $ \langle p_{k-1}, \omega_{k-1} \rangle \vDash_{\lambda_f} \phi_2$  or there exists an abstract-successor $\langle p', \omega' \rangle$ s.t. ($ \langle p_{k-1}, \omega_{k-1} \rangle \vDash_{\lambda_f}  \phi_{1} $  and $ \langle p', \omega' \rangle \vDash_{\lambda_f}  E[\phi_{1} U^a \phi_{2}]$). Since $\phi_2$ should eventually hold, $\phi_2$ should hold at $\langle p_{k-1}, \omega_{k-1} \rangle$ because the abstract-successor of $\langle p_{k-1}, \omega_{k-1} \rangle$  on this abstract-path is $\bot$. To ensure this, we move $\langle p_{k-1}, \omega_{k-1} \rangle$ to the trap configuration $\langle p_{\bot}, \gamma_{\bot} \rangle$ and add a loop here by the transition rule \ref{BCARET_item:standRulesLoopToTrapAbstractOfReturn}. In addition, we don't set $p_{\bot}$  as an element of the B\"{u}chi accepting condition set, which means that  $\langle p_{k-1}, \omega_{k-1} \rangle \vDash_{\lambda_f}   E[\phi_{1} U^a \phi_{2}]$ iff $\langle p_{k-1}, \omega_{k-1} \rangle \vDash_{\lambda_f}  \phi_2$ by the transition rules in \ref{BCARET_item:standRulesEUA}. This ensures the liveness requirement $\phi_2$ in $E [\phi_1 U^a \phi_2]$ is eventually satisfied.}

						\item {With regards to finite abstract paths $\langle p_0, \omega_0 \rangle \langle p_1, \omega_1 \rangle... \langle p_m, \omega_m \rangle$ where $\langle p_{m}, \omega_{m} \rangle \allowbreak \xRightarrow{}_{\mathcal{P}} \langle p_{m+1}, \omega_{m+1} \rangle$ corresponds to a call statement but this call never reaches its corresponding return-point, the fact that the liveness requirement $\phi_{2}$ in $E[\phi_{1} U^g \phi_{2}]$ is eventually satisfied in $\mathcal{P}$ is ensured by the fact that $p \notin F$.  Look at Figure \ref{BCARET_fig:nextAbstractWithoutMatching} where the procedure $proc$ never terminates. In this figure, for every $0 \leq u \leq i$, the abstract path starting from $\langle p_u, \omega_u \rangle$ is finite. Suppose that we want to check whether  $ \langle p_i, \omega_i \rangle \vDash_{\lambda_f}  E[\phi_{1} U^a \phi_{2}]$, then, we get that $ \langle p_i, \omega_i \rangle \vDash_{\lambda_f}  E[\phi_{1} U^a \phi_{2}]$ iff $ \langle p_{i}, \omega_{i} \rangle \vDash_{\lambda_f} \phi_2$  or there exists an abstract-successor $\langle p', \omega' \rangle$ s.t. ($ \langle p_{i}, \omega_{i} \rangle \vDash_{\lambda_f}  \phi_{1} $  and $ \langle p', \omega' \rangle \vDash_{\lambda_f}  E[\phi_{1} U^a \phi_{2}]$). Since $\phi_2$ should eventually hold, $\phi_2$ should hold at $\langle p_{i}, \omega_{i} \rangle$ because the abstract-successor of $\langle p_{i}, \omega_{i} \rangle$  on this abstract-path is $\bot$.  As explained above, at $\langle p_{i}, \omega_{i} \rangle$, we will encode the formula $E[\phi_{1} U^a \phi_{2}]$ into the stack and mimic the run of $\mathcal{P}$ on $\mathcal{BP}_\varphi$ until it reaches the corresponding return-point of the call. In other words, if the call is never reached, the run of $\mathcal{BP}_\varphi$ will infinitely visit the control locations of $\mathcal{P}$. To ensure this path unaccepted, we don't set $p \in P$ as an element of the B\"{u}chi accepting condition set, which means that  $\langle p_{i}, \omega_{i} \rangle \vDash_{\lambda_f}   E[\phi_{1} U^a \phi_{2}]$ iff $\langle p_{i}, \omega_{i} \rangle \vDash_{\lambda_f}  \phi_2$ by the transition rules in \ref{BCARET_item:standRulesEUA}. This ensures the liveness requirement $\phi_2$ in $E [\phi_1 U^a \phi_2]$ is eventually satisfied.

						}

					\end{itemize}

				\end{mycolorforparablue}


				\medskip
				\noindent
				\changenaas{Thus, we can show that:}

				
				\begin{theorem}
					\label{BCARET_theorem:BCARETWitStandardValuation}
					Given a PDS  $\mathcal{P} = (P, \Gamma, \Delta, \sharp)$, a set of atomic propositions $AP$, a labelling function $f: AP \rightarrow 2^P$ and a BCARET formula  $\varphi$, we can compute an ABPDS $\mathcal{BP}_\varphi$ such that for every configuration $\langle p, \omega \rangle$, $\changeallfullversiona{\langle p, \omega \rangle} \changenaad{\vDash_{\lambda_f}} \varphi$ iff $\mathcal{BP}_\varphi$ has an accepting run from the configuration $\langle \changenaad{\llparenthesis p, \varphi \rrparenthesis}, \omega \rangle$
					
				\end{theorem}

				%
				%
				%
				%

				\notesctl{I just changed complexity, the number of transitions from  $\mathcal{O}((|P||\Gamma| + |\Delta|) |\varphi|)$ to \changenaar{$\mathcal{O}(|P||\Gamma| |\Delta| |\varphi|)$}. Please help me to re-check this.}
				
				\noindent
				\changenaal{The number of control locations of $\mathcal{BP}_\varphi$ is at most $\mathcal{O}(|P||\varphi|)$, \changenaas{the number of stack symbols is at most $\mathcal{O}(|\Gamma||\varphi|)$} and the number of transitions is at most \changenaas{$\mathcal{O}(|P||\Gamma| |\Delta| |\varphi|)$}. Therefore, \changenaas{we get from Theorems \ref{BCARET_theorem:ABPDSMembershipProblem} and \ref{BCARET_theorem:BCARETWitStandardValuation}:}}

				\begin{theorem}
					\changenaal{Given a PDS  $\mathcal{P} = (P, \Gamma, \Delta, \sharp)$, a set of atomic propositions $AP$, a labelling function $f: AP \rightarrow 2^P$ and a BCARET formula  $\varphi$, for every configuration $\langle p, \omega \rangle \in P \times \Gamma^*$, whether or not $\langle p, \omega \rangle$ satisfies $\varphi$ can be solved in time  \changenaas{$\mathcal{O}(|P|^2   |\varphi|^3. |\Gamma|(|P| |\Gamma| |\Delta|. |\varphi|. 2^{5 |P| |\varphi|} + 2^{|P| |\varphi|}. |\omega|)) $ } }
					
				\end{theorem}

				\notesctl{This is the sum of the complexity to compute the Alternating Multi AUtomata s.t. $\changeallfullversiona{\langle p, \omega \rangle} \vDash_{\lambda} \varphi$  AND the time to check the membership on AMA. }

				\section{BCARET model-checking for PDSs with regular valuations}
				
				Up to now, we have considered the \textit{standard} model-checking problem for BCARET, where the validity of an atomic proposition depends only on the control state, not on the stack.  In this section, we go further and consider \changenaaw{model-checking with regular valuations} where the set of configurations in which an atomic proposition holds is a regular set of configurations \changenaav{(see \changenaaw{Section \ref{BCARET_sec:BCARETDefinition}} for a formal definition of \changenaau{regular valuations})}.

				%
				%
				%
				%
				%

				\subsection{From BCARET model checking of PDSs with regular valuations to the membership problem in ABPDSs}

				\notesctl{You want me to define a new BCARET with regular valuations? I think it is the same with the definition of BCARET. Regular Valuations is just the way to validate an atomic proposition. In definition of BCARET, we already use $\lambda$ instead of labelling function. So, it already include this case?? WHat do you think??. }


				Given a pushdown system $\mathcal{P} = (P, \Gamma, \Delta, \sharp)$, \changenaas{and} a set of \changenaal{atomic propositions $AP$}, \changenaas{let} $\varphi$ be a \changenaal{BCARET} formula over \changenaad{$AP$,}  $\lambda: AP \rightarrow 2^{P \times \Gamma^*}$ be a labelling function s.t. for every $e \in AP$, $\lambda(e)$ is a regular set of configurations. Given a configuration $c_0$, we propose in this section an algorithm to \changenaal{check} whether $\changeallfullversiona{c_0} \vDash_{\lambda} \varphi$. Intuitively, we \changenaal{compute} an ABPDS $\mathcal{BP}'_\varphi$ s.t. $\mathcal{BP}'_\varphi$ recognizes a configuration $c$ of $\mathcal{P}$ iff $\changeallfullversiona{c} \vDash_{\lambda} \varphi$. Then, to check if $\changeallfullversiona{c_0}$ satisfies $\varphi$, \changenaal{we will} check whether $\mathcal{BP}'_\varphi$ recognizes $c_0$.

				
				For every $e \in AP$, since $\lambda(e)$ is a regular set of configurations,  let $M_e = (Q_e, \Gamma, \delta_e, I_e, F_e)$ be a multi-automaton s.t. $L(M_e) = \lambda(e)$,   $M_{\neg e} = (Q_{\neg e}, \Gamma, \delta_{\neg e}, I_{\neg e}, F_{\neg e})$ be a multi-automaton s.t.  $L(M_{\neg e}) = P \times \Gamma^* \setminus \lambda(e)$, which means $M_{\neg e}$ \changenaat{will recognize} the complement of $\lambda(e)$ that is the set of configurations \changenaas{in which $e$} doesn't hold. Note that for every $e \in AP$, the \changenaat{initial states} of $M_e$ and $M_{\neg e}$ are the control locations $p \in \mathcal{P}$.  Thus, to distinguish between the initial states of these two automata, we will denote the initial state corresponding to the control location $p$ in $M_e$ (resp. $M_{\neg e}$) by $p_e$ (resp. $p_{\neg e}$). \changenaas{Let $AP^+(\varphi) = \{ e \in AP \;|\; e \in Cl(\varphi)\}$ and $AP^-(\varphi) = \{ e \in AP \;|\; \neg e \in Cl(\varphi)\}$.}

				Let $\mathcal{BP}'_\varphi = (P'', \Gamma'', \Delta'', F')$ be the ABPDS defined as follows:
				
				\ifdefined \NotNeedToReducePages \begin{itemize} \else \begin{itemize}[noitemsep,topsep=0pt] \fi 
						\item {\changenaal{$P'' =  P \cup P \times Cl(\varphi) \changeallfullversionb{\cup \{ p_{\bot} \}} \cup \bigcup_{e \in AP^+ (\varphi)} Q_e \cup \bigcup_{e \in AP^- (\varphi)} Q_{\neg e}$}}
						
						\item {$\Gamma'' = \Gamma \cup (\Gamma \times Cl(\varphi)) \changeallfullversionb{\cup \{ \gamma_{\bot} \}}$}
						\item {$F' = F_1 \cup F_2 \cup F_3$ where 
							
							\ifdefined \NotNeedToReducePages \begin{itemize} \else \begin{itemize}[noitemsep,topsep=0pt] \fi 
									\item {\changenaao{$F_1 = \bigcup_{e \in AP^+ (\varphi)} F_e $}}
									\item {\changenaao{$F_2 =  \bigcup_{e \in AP^- (\varphi)} F_{\neg e} $}}
									\item {$F_3 = \{P \times Cl_{R} (\varphi) \} $ where $Cl_{R} (\varphi)$ is the set of formulas of $Cl(\varphi)$ \changenaal{in the form} $E [\varphi_1 R^b \varphi_2]$ or $A [\varphi_1 R^b \varphi_2]$ \changenaab{($b \in \{g, a\}$)}  }
								\end{itemize}
							}
							
						\end{itemize}
						
						The transition relation $\Delta''$ is the smallest set of transition rules defined as follows: \changenaas{$\Delta \subseteq \Delta''$}, \changenaas{$\Delta'_0 \subseteq \Delta''$ where $\Delta'_0$ is the transitions of $\Delta'$ that are created by the rules \changeallfullversionb{from \changenaasbb{\ref{BCARET_item:standRulesAnd}} to \ref{BCARET_item:standRulesLoopToTrapAbstractOfReturn}}  and} \changenaat{such that:}

						\begin{enumerate}[label=($\beta$\arabic*) ]
							
							\item {\changenaaw{for every $p \in P$, $e \in AP^+ (\varphi)$, $\gamma \in \Gamma$}:  $\langle \llparenthesis p, \changenaaw{e} \rrparenthesis, \gamma \rangle \rightarrow \langle  \changenaaw{p_e}, \gamma \rangle \in \Delta''$} \label{BCARET_item:regularRulesAP}
							
							\item {\changenaaw{for every $p \in P$, $e \in AP^- (\varphi)$, $\gamma \in \Gamma$}: $\langle \llparenthesis p, \changenaaw{\neg e} \rrparenthesis, \gamma \rangle \rightarrow \langle \changenaaw{ p_{\neg e}}, \gamma \rangle \in \Delta''$} \label{BCARET_item:regularRulesNegAP}

							\item {\label{BCARET_item:standRulesRunOnMultiAutomaton} for very $\changenaae{(q_1, \gamma, q_2)   \in } (\bigcup_{e \in AP^+ (\varphi)} \delta_e) \cup (\bigcup_{e \in AP^- (\varphi)} \delta_{\neg e})$: $\langle q_1, \gamma \rangle \rightarrow \langle q_2, \epsilon \rangle \in \changenaal{\Delta''}$}

							\item {\label{BCARET_item:standRulesAddLoopOnBottomSymbol} for very  $ q \in (\bigcup_{e \in AP^+ (\varphi)} F_e) \cup (\bigcup_{e \in AP^- (\varphi)} F_{\neg e})$: $\langle q, \sharp \rangle \rightarrow \langle q, \sharp \rangle \in \changenaal{\Delta''}$
							}
							
						\end{enumerate}

						%
						%
						%
						%
						%
						%
						%
						%
						%
						

						Intuitively, we compute the ABPDS $\mathcal{BP}'_\varphi$ such that $\mathcal{BP}'_\varphi$ has an accepting run from $\langle \llparenthesis p, \phi \rrparenthesis, \omega \rangle$ iff the configuration $\langle p, \omega \rangle$ satisfies $\phi$ according to the \changenaas{regular labellings $M_e$} for every $e \in AP$. \changenaas{The only difference with the previous case of standard \changenaat{valuations}, where an atomic proposition holds at a configuration depends only on the control location of that configuration, not on its stack, comes from the interpretation of the atomic proposition $e$. This is why \changenaat{$ \Delta''$ contains $\Delta$ and $\Delta'_0$ (which are the transitions of $\mathcal{BP}_\varphi$ that don't consider the atomic propositions). Here the rules $(\beta_1)-(\beta_4)$ deal with} the cases \changenaaw{$e$, $\neg e$ ($e \in AP$)}.} Given $p \in P$, $\phi = e \in AP$, $\omega \in \Gamma^*$, we get that the ABPDS  $\mathcal{BP}'_\varphi$ should accept \changenaao{$\langle \llparenthesis p, e \rrparenthesis, \omega \rangle$} iff $\langle p, \omega \rangle \in L(M_e)$. To check whether $\langle p, \omega \rangle \in L(M_e)$, we let $\mathcal{BP}'_\varphi$ go to state $p_e$, the initial state corresponding to $p$ in $M_e$ by adding rules in \changeni{\changenaasbb{\ref{BCARET_item:regularRulesAP}}}; and then, from this state, we will check whether $\omega$ is accepted by $M_e$. This is ensured by the transition rules in \changeni{\changenaasbb{\ref{BCARET_item:standRulesRunOnMultiAutomaton}}} and \changeni{\changenaasbb{\ref{BCARET_item:standRulesAddLoopOnBottomSymbol}}}. \changeni{\changenaasbb{\ref{BCARET_item:standRulesRunOnMultiAutomaton}}} lets $\mathcal{BP}'_\varphi$ mimic a run of $M_e$ on $\omega$, i.e., if $\mathcal{BP}'_\varphi$  is in a state $q_1$ with $\gamma$ on the top of the stack, and if \changenaaw{$(q_1, \gamma, q_2)$} is a transition rule in $M_e$, then, $\mathcal{BP}'_\varphi$  will move to state $q_2$ and pop $\gamma$ from its stack. Note that popping $\gamma$ allows us to check the rest of the word. In $M_e$, a configuration is accepted if the run with the word $\omega$ \changenaal{reaches} the final state in $F_e$; i.e., if $\mathcal{BP}'_\varphi$ reaches a state $q \in F_e$ with an empty stack, i.e., with a stack containing the bottom stack symbol $\sharp$. Thus, we add $F_e$ as a set of accepting control locations in $\mathcal{BP}'_\varphi$. \changenaas{Since} $\mathcal{BP}'_\varphi$  only recognizes infinite paths, \changenaas{\ref{BCARET_item:standRulesAddLoopOnBottomSymbol} adds a loop on every configuration $\langle q, \sharp \rangle$ where $q \in F_e$}. The intuition \changenaas{behind the transition rules in} \changeni{\changenaasbb{\ref{BCARET_item:regularRulesNegAP}}} is similar \changenaas{to that of \changenaasbb{\ref{BCARET_item:regularRulesAP}}}. \changenaas{They correspond to the case where}  $\phi = \neg e$.


						\begin{theorem}
							\label{BCARET_theorem:BCARETWithRegularValuation}
							Given a PDS  $\mathcal{P} = (P, \Gamma, \Delta, \sharp)$, a set of atomic propositions $AP$, \changenaab{a \textit{regular} labelling function $\lambda: AP \rightarrow 2^{P \times \Gamma^*}$} and a BCARET formula  $\varphi$, we can compute an ABPDS $\mathcal{BP}'_\varphi$ such that for every configuration $\langle p, \omega \rangle$, $\changeallfullversiona{\langle p, \omega \rangle} \vDash_{\lambda} \varphi$ iff $\mathcal{BP}'_\varphi$ has an accepting run from the configuration $\langle \changenaad{\llparenthesis p, \varphi \rrparenthesis}, \omega \rangle$
						\end{theorem}
						

						%
						%
						%
						%

						\noindent
						\changenaas{The number of control locations of $\mathcal{BP}'_\varphi$ is at most $\mathcal{O}(|P||\varphi| + k)$ where $k = \sum_{e \in AP^+(\varphi) } |Q_e| + \sum_{e \in AP^{-}(\varphi) } |Q_{\neg e}|$, \changenaas{the number of stack symbols is at most $\mathcal{O}(|\Gamma||\varphi|)$} and the number of transitions is at most \changenaas{$\mathcal{O}(|P||\Gamma| |\Delta| |\varphi| + d)$}  where $d = \sum_{e \in AP^+(\varphi) } |\delta_e| + \sum_{e \in AP^{-}(\varphi) } |\delta_{\neg e}|$. Therefore, \changenaas{we get from Theorems \ref{BCARET_theorem:ABPDSMembershipProblem} and \ref{BCARET_theorem:BCARETWithRegularValuation}:}}

						\begin{theorem}
							\changenaas{Given a PDS  $\mathcal{P} = (P, \Gamma, \Delta, \sharp)$, a set of atomic propositions $AP$, \changenaas{a \textit{regular} labelling function $\lambda: AP \rightarrow 2^{P \times \Gamma^*}$} and a BCARET formula  $\varphi$, for every configuration $\langle p, \omega \rangle \in P \times \Gamma^*$, whether or not $\langle p, \omega \rangle$ satisfies $\varphi$ can be solved in time \changenaas{$\mathcal{O}((|P||\varphi| + k)^2. |\Gamma| |\varphi| ((|P||\Gamma| |\Delta| |\varphi| + d). 2^{5 (|P| |\varphi| + k)} + 2^{|P| |\varphi| + k}. |\omega|)) $ } }
							
						\end{theorem}

						\section{Conclusion}
						\label{BCARET_sec:conclusion}

						In this \changeifma{paper}, we introduce the Branching temporal logic of CAlls and RETurns BCARET and show how it can be used to describe malicious behaviors that CARET and other specification formalisms cannot. We present an algorithm for "standard" BCARET model checking for PDSs where whether a configuration of a PDS satisfies an atomic proposition or not depends only on the control location of that configuration. Moreover, we consider BCARET model-checking for PDSs with regular valuations where the set of configurations on which an atomic proposition holds is a regular language. Our approach is based on reducing these problems to the emptiness problem of Alternating B\"{u}chi Pushdown Systems.


						
						
						\newpage
						\bibliographystyle{plain} 

						\ifdefined \UseFullVersion  
						\newpage
						\input{proof_BCARET.tex}
						\else

						\fi

					\end{document}